\documentclass[sn-mathphys]{sn-jnl}% Math and Physical Sciences Reference Style
\usepackage{upgreek}
\jyear{2022}%

\theoremstyle{thmstyleone}%
\theoremstyle{thmstyletwo}%

\theoremstyle{thmstylethree}%

\begin{document}
\title[Article Title]{The 100-m X-ray Test Facility at IHEP}


\author*[1]{\fnm{Yusa} \sur{Wang}}\email{wangyusa@ihep.ac.cn}
\author[1]{\fnm{Zijian} \sur{Zhao}}\email{zhaozijian@ihep.ac.cn}
\author[1]{\fnm{Dongjie} \sur{Hou}}\email{houdj@ihep.ac.cn}
\author[1]{\fnm{Xiongtao} \sur{Yang}}\email{yangxt@ihep.ac.cn}
\author[1]{\fnm{Can} \sur{Chen}}\email{chencan@ihep.ac.cn}
\author[1]{\fnm{Xinqiao} \sur{Li}}\email{lixq@ihep.ac.cn}
\author[1]{\fnm{Yuxuan} \sur{Zhu}}\email{zhuyx@ihep.ac.cn}
\author[1]{\fnm{Xiaofan} \sur{Zhao}}\email{zhaoxf@ihep.ac.cn}
\author[1]{\fnm{Jia} \sur{Ma}}\email{majia@ihep.ac.cn}
\author[1]{\fnm{He} \sur{Xu}}\email{xuhe@ihep.ac.cn}
\author[1]{\fnm{Yupeng} \sur{Chen}}\email{chenyp@ihep.ac.cn}
\author[1]{\fnm{Guofeng} \sur{Wang}}\email{wanggf@ihep.ac.cn}
\author[1]{\fnm{Fangjun} \sur{Lu}}\email{lufj@ihep.ac.cn}
\author[1]{\fnm{Shuangnan} \sur{Zhang}}\email{zhangsn@ihep.ac.cn}
\author[1]{\fnm{Shu} \sur{Zhang}}\email{szhang@ihep.ac.cn}
\author[1]{\fnm{Yong} \sur{Chen}}\email{ychen@ihep.ac.cn}
\author[1]{\fnm{Yupeng} \sur{Xu}}\email{xuyp@ihep.ac.cn}
\affil*[1]{\orgdiv{Key Laboratory for Particle Astropysics}, \orgname{Institute of High Energy Physics, Chinese Academy of Sciences}, \orgaddress{\street{Yuquan Road}, \city{Beijing}, \postcode{100049}, \country{China}}}


\abstract{The 100-m X-ray Test Facility of the Institute of High Energy Physics (IHEP) was initially proposed in 2012 for the test and calibration of the X-ray detectors of the Hard X-ray Modulation Telescope (HXMT) with the capability to support future X-ray missions. The large instrument chamber connected with a long vacuum tube can accommodate the X-ray mirror, focal plane detector and other instruments. The X-ray sources are installed at the other end of the vacuum tube with a distance of 105 m, which can provide an almost parallel X-ray beam covering 0.2$\sim$60 keV energy band. The X-ray mirror modules of the Einstein Probe (EP) and the enhanced X-ray Timing and Polarimetry mission (eXTP) and payload of the Gravitational wave high-energy Electromagnetic Counterpart All-sky Monitor (GECAM) have been tested and calibrated with this facility. It has been also used to characterize the focal plane camera and aluminum filter used on the Einstein Probe. In this paper, we will introduce the overall configuration and capability of the facility, and give a brief introduction of some calibration results performed with this facility.}

%%================================%%
%% Sample for structured abstract %%
%%================================%%

\keywords{Test facility, X-ray astronomy, X-ray telescopes, X-ray optics, Calibration}



\maketitle

\section{Introduction}\label{sec1}
\begin{figure}[!htp]
\centering
\includegraphics[width=0.9\textwidth]{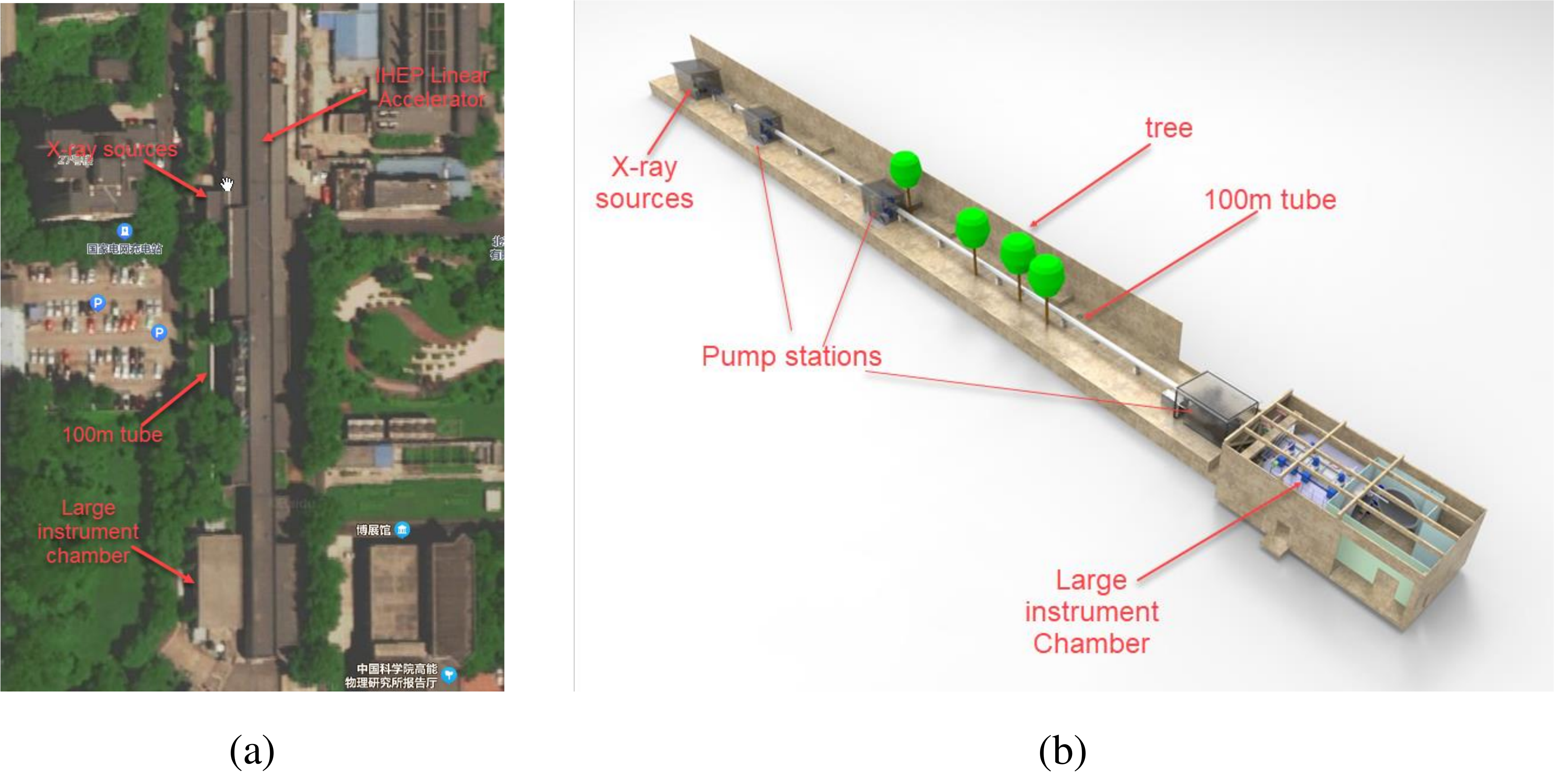}
\caption{(a): The location of the 100XF on the campus of IHEP, (b): The overall structure of 100XF.}
\label{fig1}
\end{figure}

The 100-m X-ray Test Facility of the Institute of High Energy Physics (100XF), as shown in Figure~\ref{fig1}, is located on the campus of the institute, in parallel to the linear accelerator of the Beijing Electron-Positron Collider. 100XF is the largest test and calibration facility in China dedicated to space astronomy mission. It was originally constructed for the calibration of the X-ray detectors of China's first X-ray astronomy satellite--the Hard X-ray Modulation Telescope (HXMT)\cite{2007HXMT}\cite{2020HXMT}, but with the capability to be upgraded for the calibration of large focusing X-ray telescopes. During its construction, IHEP was also involved in developing several other X-ray astronomy missions, such as the Einstein Probe (EP)\cite{2018EP}, the Gravitational wave high-energy Electromagnetic Counterpart All-sky Monitor (GECAM)\cite{GECAM} and the enhanced X-ray Timing and Polarimetry mission (eXTP)\cite{2019eXTP}. Some of the instruments of these satellites have already been tested or calibrated on 100XF.

The calibration of X-ray telescopes is a basic and crucial procedure in the development of an X-ray mission. At the beginning of the development, the X-ray optics and detectors need to be tested iterately for verification of designs. The performance of the whole detector/telescope assemblies needs to be checked and calibrated in detail before launch, for complete understanding of the instruments. Even when an instrument works in orbit, there are essential ground calibration contents for deeper understanding of the instrument, which are useful for in-orbit calibration. Only with these test and calibration data, the analyses of scientific data will be credible. Especially, the test of large X-ray optics is expensive and demanding for X-ray facility. For the previous X-ray missions, several X-ray facilities were built for ground calibration. Max-Planck-Institut f\"{u}r extraterrestrische Physik (MPE) built the PANTER X-ray facility\cite{panter1979}\cite{panter} for ROSAT mission, and played an important role in XMM-Newton\cite{XMM} and eROSITA\cite{eROSITA}; its vacuum tube is 130 m long. NASA built the Marshall Space Flight Center’s X-ray Calibration Facility (XRCF)\cite{XRCF} for the test and calibration of the Chandra mission\cite{Chandra}; its vacuum tube is 500 m long. 

In addition, some other similar calibration facilities are in use ,e.g.: the X-ray Astronomy Calibration and Testing (XACT) facility of Istituto Nazionale di AstroFisica (INAF) located in Italy\cite{XACT}, the The Leicester Long Beam-line Test Facility (LLBTF) at the University of Leicester in UK, the X-ray facility of the Space Research Institute of the Russian Academy of Sciences (IKI), The NASA Marshall Space Flight Center (MSFC) 100-m X-ray Beamline (also known as the Stray Light Test Facility, SLTF). Some other faclities such as the Beam Ex pander Testing X-ray (BEaTriX) of INAF\cite{BEaTriX} and Vertical X-ray raster-scan facility (VERT-X)\cite{vert-x}, are under construction or proposed to satisfy the future missions' requirements. All the facilities mentioned above are well introduced by Bianca Salmaso in his paper soon to be published. It must be noted that the minimal length of the X-ray vacuum tube for Wolter-I (diameter is less than 300 mm) with reasonable angular resolution should be more than 75 m long.

The 100XF has a 100 m vacuum tube, three pump stations and a large instrument chamber in which the instruments can be installed and tested. In order to meet various requirements of test and calibrations, the facility adopts several X-ray sources and detectors. For example, a modulated X-ray source is applied to verify the timing performance of the tested device (focal plane detectors and whole telescopes),  and a large format sCMOS X-ray camera is installed for the imaging of a large area.

This paper gives a detailed description of the facility and its key performances. Some results of the characterization of focal plane camera and mirror module of the Follow-up X-ray Telescope (FXT) of EP as well as of the X-ray mirror shells of eXTP, which have been tested in this facility, are also discussed in this paper.

\section{Design of 100XF}\label{sec2}

The 100XF is mainly composed of: the large instrument chamber (hereinafter referred to as chamber), 100 m tube, pump stations, movable stages, X-ray sources and standard X-ray detectors. The chamber can accommodate three movable stages and the instruments to be tested. Various flanges are installed on the chamber wall for data and command transmissions and cooling water (or alcohol) supply. The 100 m tube with three pump stations connects the chamber and the X-ray sources and at the same time realizes the quasi-parallel beam.

\subsection{the large instrument chamber \& 100 m tube}\label{subsec2-1}

\begin{figure}[!htp]
\centering
\includegraphics[width=0.72\textwidth]{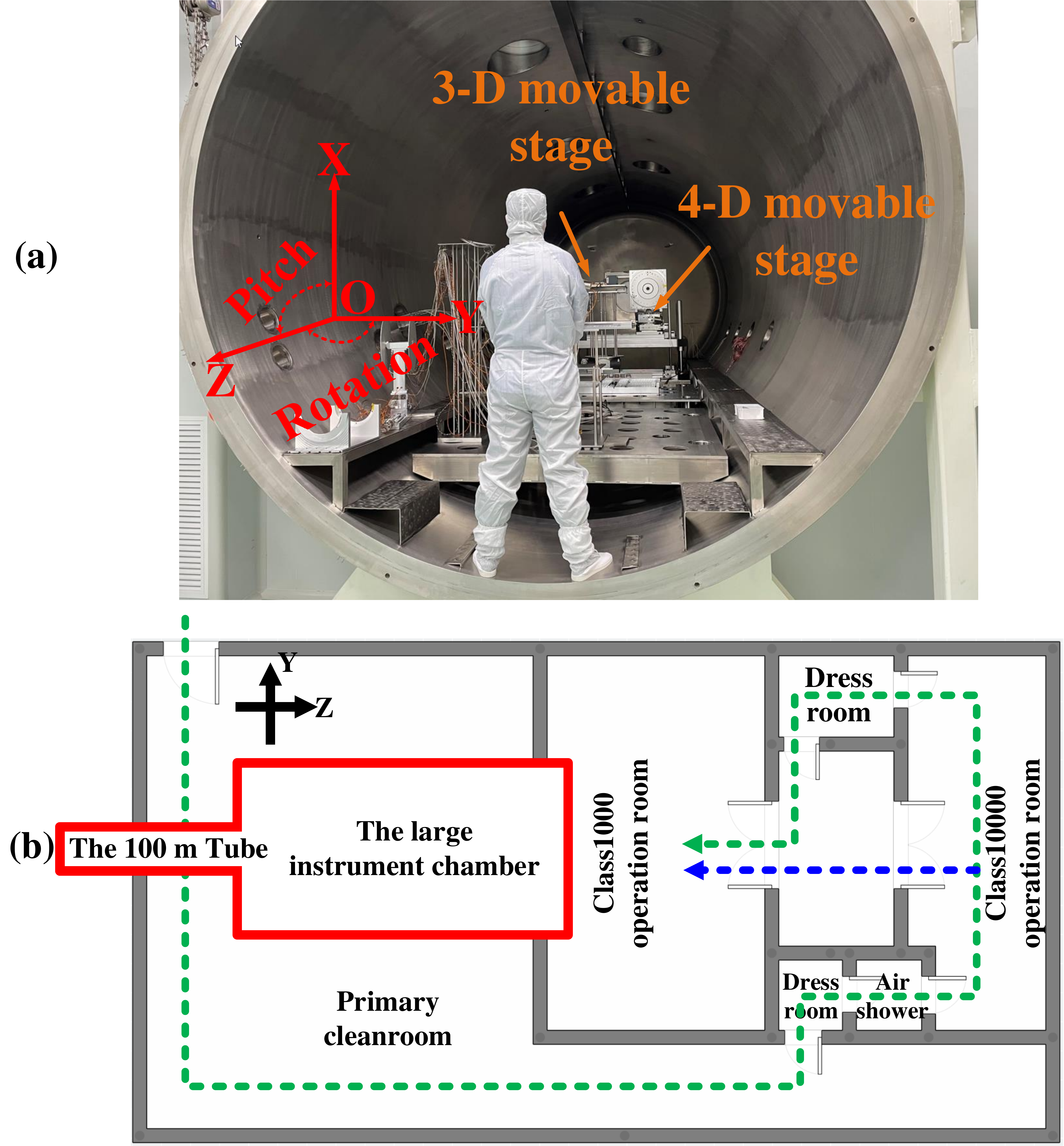}
\caption{(a) Inside the large instrument chamber. The right-hand Cartesian coordinate indicates the position coordinates of the whole system, where the pitch motion and rotation motion are the turn around the Y and X axis, arespectively. The relative positions of the 4-D and 3-D movable stages are shown by the orange arrows. The the man is roughly at the position of the 6-D stage.  There are 4-D, 3-D and 6-D movable stages, respectively, from inside to outside of the chamber. (b) Schematic diagram of the chamber and clean environment configuration. The black coordinate in the left upper corner corresponds to the coordinate in Figure(a), the green and blue dotted arrows show the routes entering the chamber operation room for operators and large instruments, respectively.}
\label{fig2}
\end{figure}

The chamber, which is 8 m long and 3.4 m in diameter\cite{FXT-zhao},  can easily accommodate X-ray telescopes such as FXT onboard EP satellite and Spectroscopy Focusing Array (SFA) onboard eXTP. As shown in Figure~\ref{fig2}(a), the inner wall of the chamber is polished (single point diamond polishing method) to keep the cleanness and high vacuum level in the chamber, so as to protect the X-ray mirrors and detectors from contamination. As Figure~\ref{fig2}(b) shows, the whole chamber is placed in a primary cleanroom except the door. The door is covered by a class-1000 (ISO7) cleanroom. The operators can reach the chamber door along the dotted green line (through the air shower and one class-10000 (ISO6) cleanroom). 

\begin{figure}[!htp]
\centering
\includegraphics[width=0.5\textwidth]{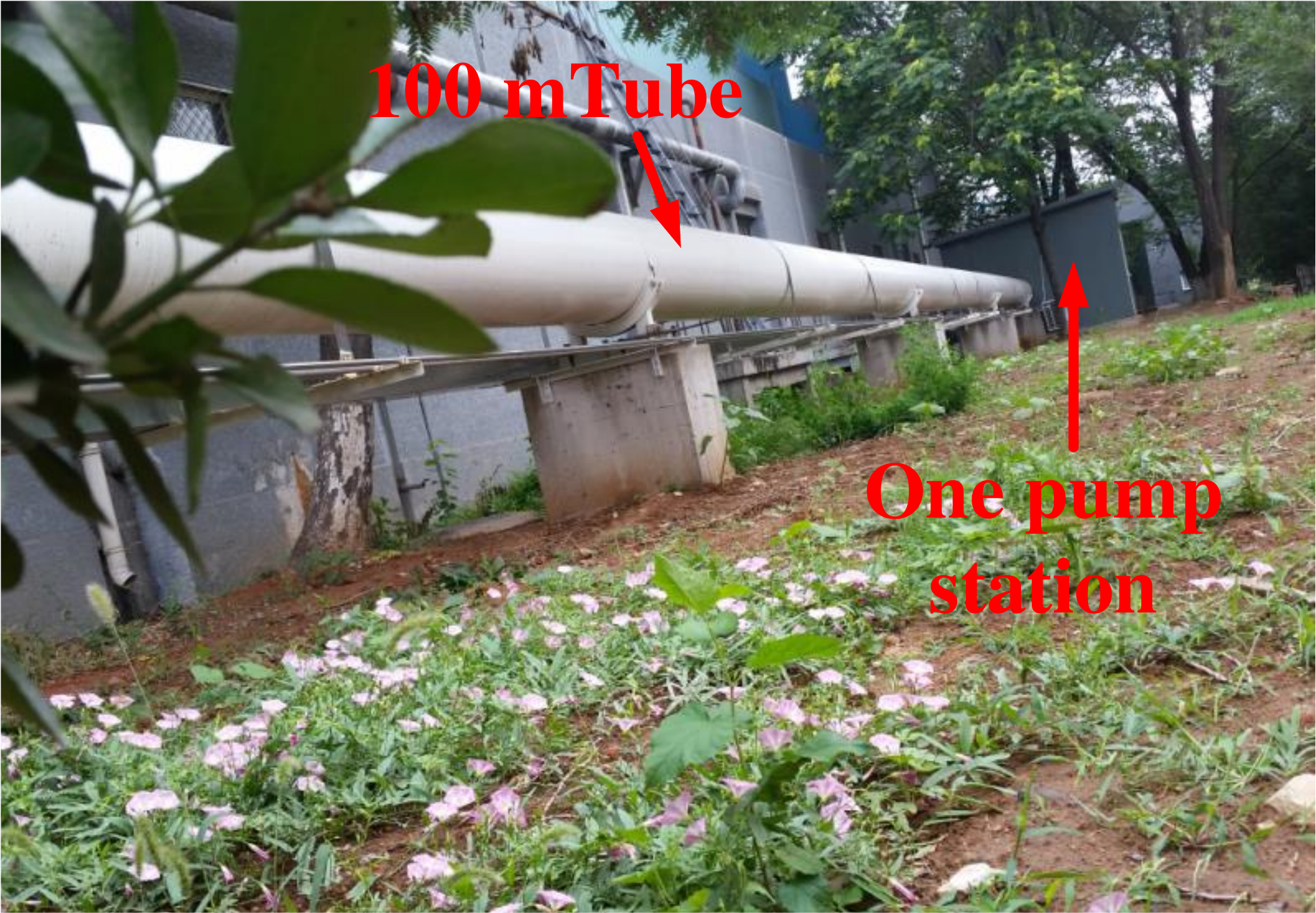}
\caption{The 100 m tube and one pump station (the middle one) of 100XF.}
\label{fig3}
\end{figure}

The 100 m (diameter 0.6 m) vacuum tube connects the chamber and X-ray sources. The X-ray beam generated by the X-ray sources passes through the tube and forms a divergent beam with a divergence angle less than $9^{\prime}$ at the exit end of the chamber under the constraint of the 100-m tube, and the divergence angle becomes smaller as the beam size reduces. There is an aperature (diameter of about 10 mm) one meter near the sources to prevent reflections of X-rays on the tube walls. 

The pump stations are used to keep the high vacuum of the whole tube, as shown in Figure~\ref{fig3}. There are 8 dry pumps (roots pump and screw pump), 7 turbo pumps (full magnetic levitation pumps) and 11 cryopumps to achieve the high vacuum efficiently. Such a combination of different pumps also makes the vacuum pumping quite fast. Within only 4 hours (1 hour for primary vacuum, 3 hours for high vacuum)the system can reach the target vacuum of 5$\times10^{-5}$ Pa and 2$\times10^{-5}$ Pa for the chamber and the tube, respectively. However, for contamination control of the chamber and test instrument, once the high vacuum is achieved, the time required to restore normal pressure exceeds 12 hours even if the chamber's vacuum is broken with moderate pressure pure nitrogen.

\subsection{Movable stages}\label{subsec2-2}

There are three movable stages inside the chamber and their relative positions are shown in Figure~\ref{fig2}. 

\begin{figure}[!htp]
\centering
\includegraphics[width=0.35\textwidth]{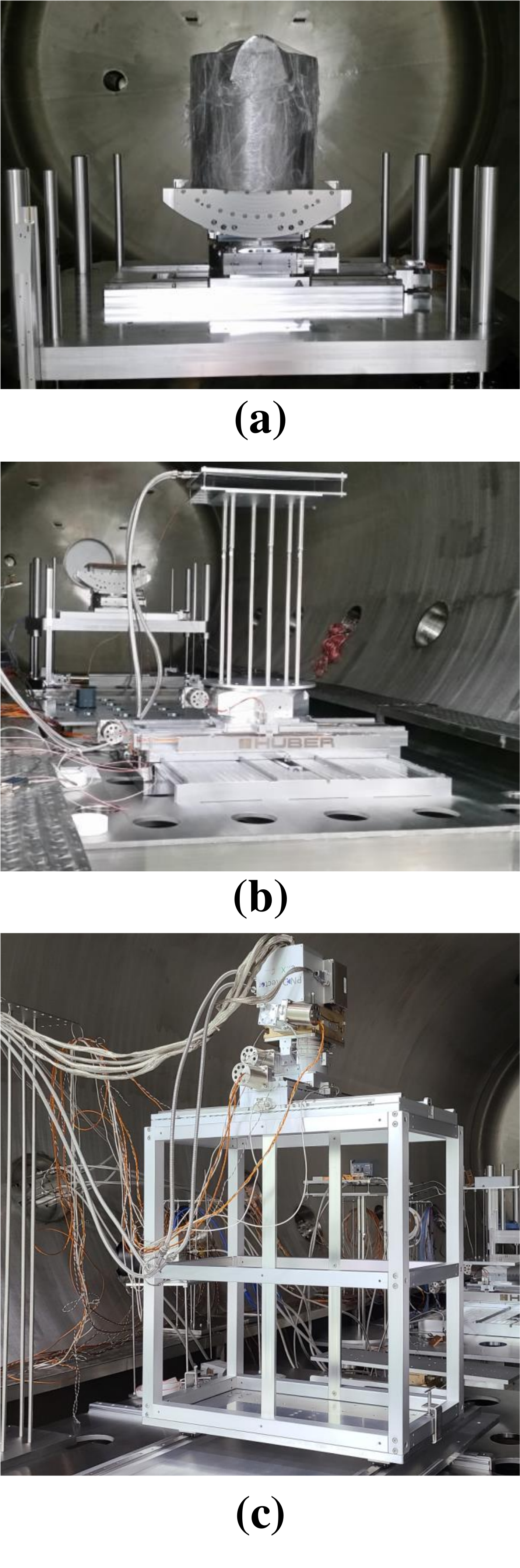}
\caption{The 4-D (a), 3-D (b), 6-D (c) movable stages in the large instrument chamber. }
\label{fig4}
\end{figure}

The innermost one is the 4-D stage (Figure~\ref{fig4}(a)), which is used to support or install the test instruments, such as optics, telescopes, payloads or even small satellites. The detailed parameters of the 4-D movable stage are listed in Table~\ref{tab1}. The load capacity of this stage is presently 200 kg but will soon be upgrated to 400 kg, so as to support the heavy instruments to be tested.

\begin{table}[!htp]
\begin{center}
\begin{minipage}{210pt}
\caption{Detailed parameters of the 4-D movable stage}\label{tab1}%
\begin{tabular}{ll}
\toprule
Name & Value\\
\midrule
Load\footnotemark[1]    & 200 kg\\
X axis range\footnotemark[2]    & 800 mm\\
X-axis repeatability    & 5 $\upmu$m\\
Z-axis range\footnotemark[2]    & 600 mm\\
Z-axis repeatability    & 10 $\upmu$m\\
Rotation range    & $360^{\circ}$\\
Rotation repeatability    & $5^{\prime \prime}$\\
Pitch range    & ${\pm}20^{\circ}$\\
Pitch repeatability    & $5^{\prime \prime}$\\
\botrule
\end{tabular}
\footnotetext[1]{The total load capacity is being upgraded to 400 kg.}
\footnotetext[2]{The coordinate directions are shown as red arrows in Figure~\ref{fig2}(a).}
\end{minipage}
\end{center}
\end{table}

The middle one is a 3-D stage (Figure~\ref{fig4}(b)) on which the focal plane cameras are installed. A cooper refrigeration plate with liquid nitrogen pipe is pre-installed on the stage to cool the detectors. Its load capacity is similar to that of the 4-D stage.

The last stage is a 6-D one (Figure~\ref{fig4}(c)). This stage is designed and assembled for the operation of X-ray cameras, detectors, especially for some X-ray telescopes with long focal length up to about 5.5 m. Its repeatability is similar to the above two stages. The combination of these three stages can meet the requirements of precision adjustment and calibration of the widely used X-ray optics and detectors, such as Wolter-I mirror modules, Micro-Pore Optics (MPO) and Micro-Slot Optics (MSO), and heavy focal plane cameras. 

\subsection{X-ray sources}\label{subsec2-3}

Properties of the X-ray sources are also critical for the test and calibration. There are several X-ray sources developed for 100XF, including a multi-target X-ray source, a double-crystal monochromator (channel cut), a modulated X-ray source and a polarized X-ray source. A specific X-ray source can be selected according to the purpose of the test or calibration. Considering the distance between the installation position of the X-ray sources and 100 m tube, the typical length between an X-ray source and mirror is about 105 m.

\begin{figure}[!htp]
\centering
\includegraphics[width=0.5\textwidth]{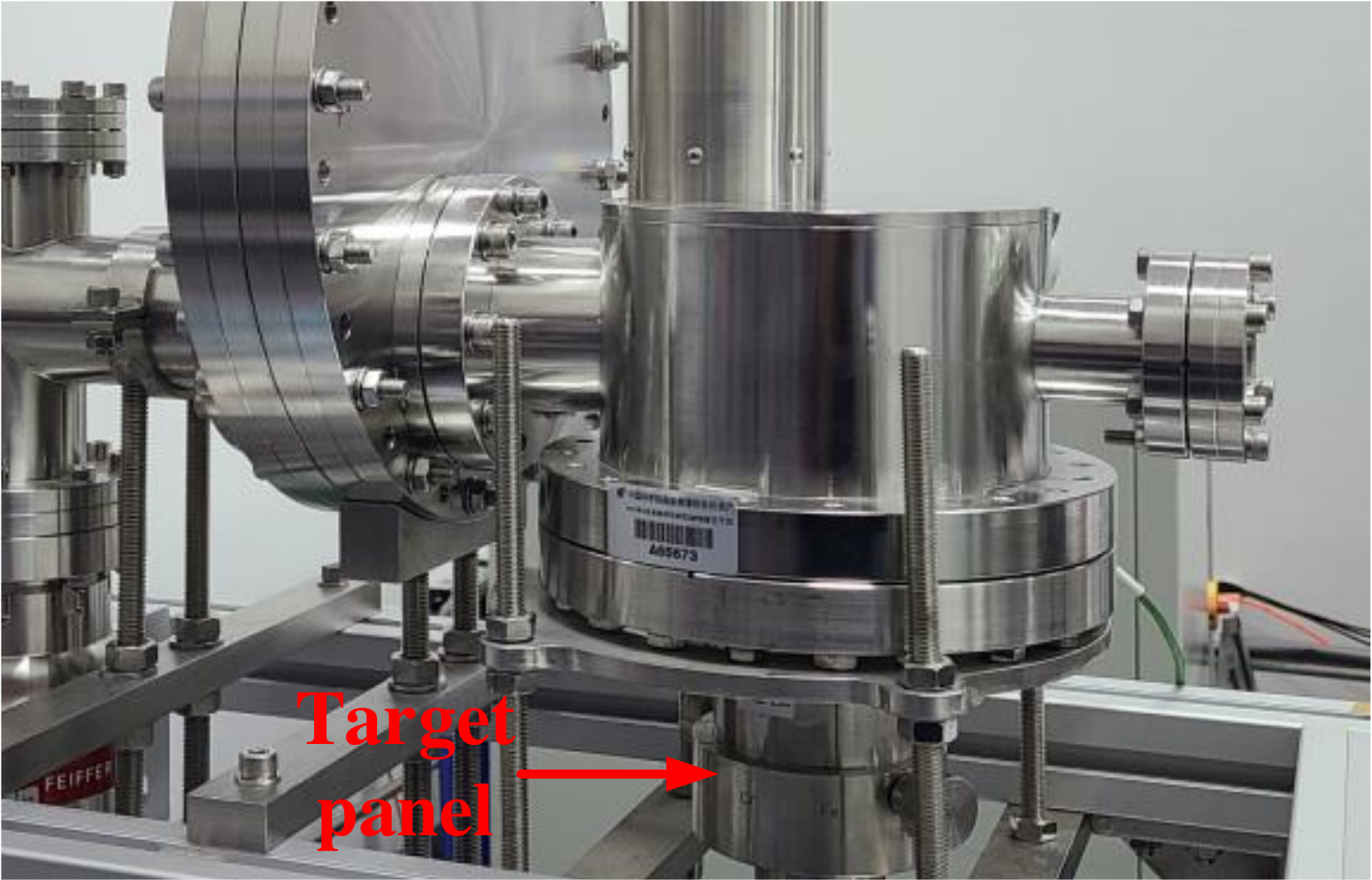}
\caption{Multi-target X-ray source. }
\label{fig5}
\end{figure}

\begin{figure}[!htp]
\centering
\includegraphics[width=0.5\textwidth]{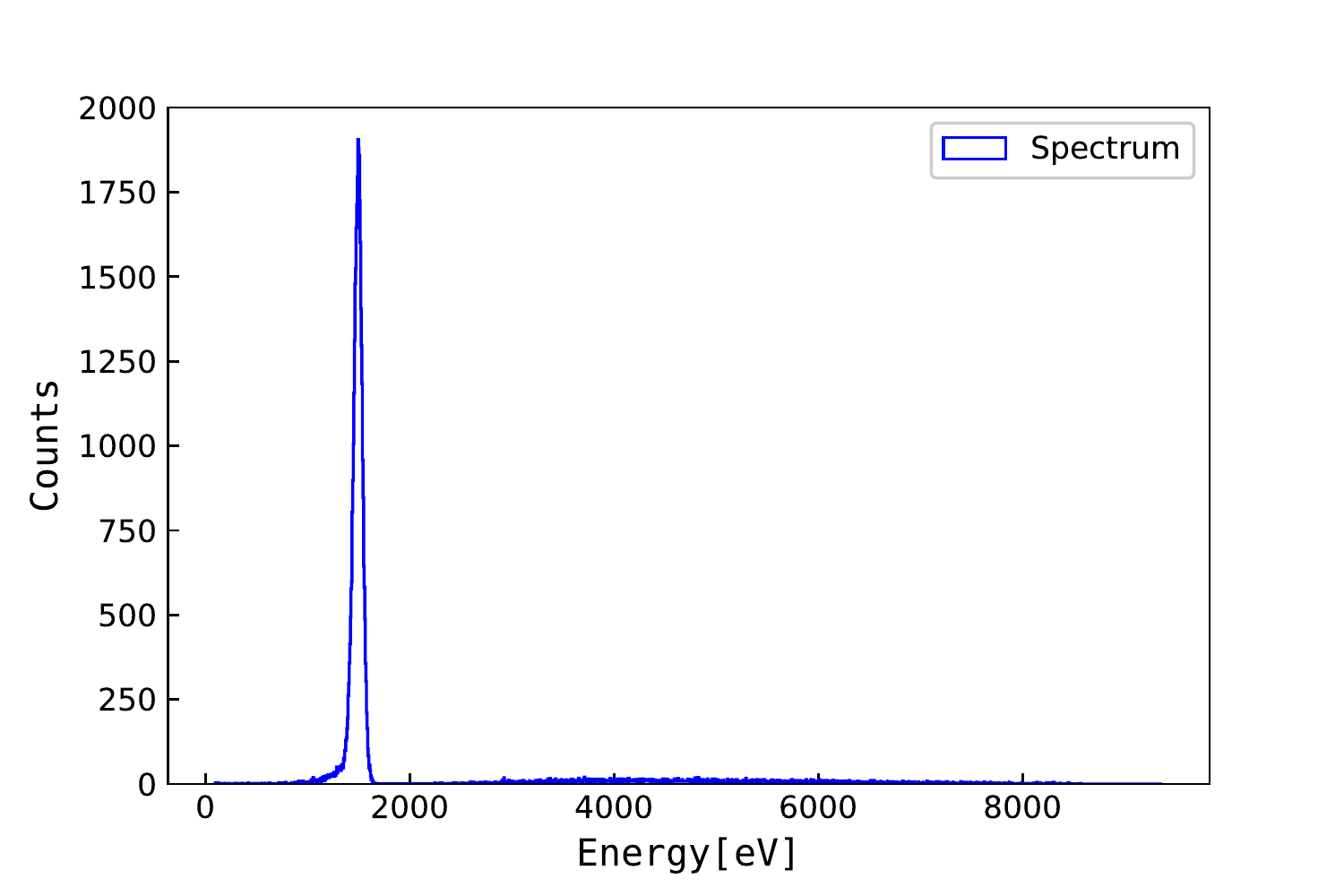}
\caption{Spectrum of aluminium target of Multi-target X-ray source, measured by pnCCD. The integration time is 200 s.}
\label{figpnspec}
\end{figure}

The multi-target electron impact X-ray source in Figure~\ref{fig5} is for basic tests of the detectors and telescopes, for example, the joint-test of the pnCCD\cite{pnccd} detector and its electronics adopted on the focal plane of EP-FXT. Figure~\ref{figpnspec} shows a spectrum of the aluminium target (with aluminium filter) of the multi-target source measured by a pnCCD. Besides aluminium, there are other targets, including carbon, magnesium, titanium, chromium, silicon, molybdenum, iron, copper and silver, with which a few fluorescent lines in 0.2$\sim$20.0 keV can be generated. All the targets can be easily changed through the knob outside the source chamber without breaking the high vacuum. In addition, 10 types of filters are mounted on the source.Some detailed information on the source are list in Table~\ref{tab2}. This source can totally meet the basic requirements of calibrating various soft X-ray telescopes. 

\begin{table}[!htp]
\begin{center}
\begin{minipage}{210pt}
\caption{Key parameters of the multi-target source}\label{tab2}%
\begin{tabular}{ll}
\toprule
Name & Value\\
\midrule
Maximum electron beam current   & 1 mA\\
Maximum voltage    & 25 kV\\
Spot size    & 0.2 mm(diameter)\\
\multirow{2}*{Flux}    & few thousand ${\rm counts/cm^{2}/s}$\\ &(exit of the 100 m tube) \\
\botrule
\end{tabular}
\end{minipage}
\end{center}
\end{table}

\begin{figure}[!htp]
\centering
\includegraphics[width=0.5\textwidth]{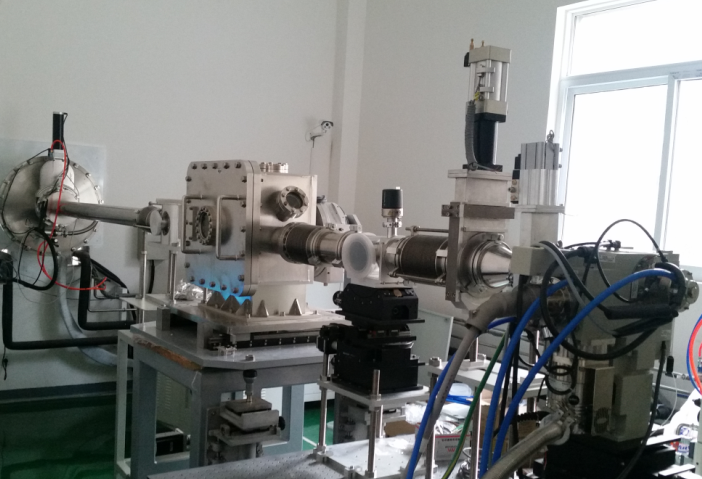}
\caption{Double-crystal monochromator (channel cut). }
\label{fig6}
\end{figure}

\begin{figure}[!htp]
\centering
\includegraphics[width=0.7\textwidth]{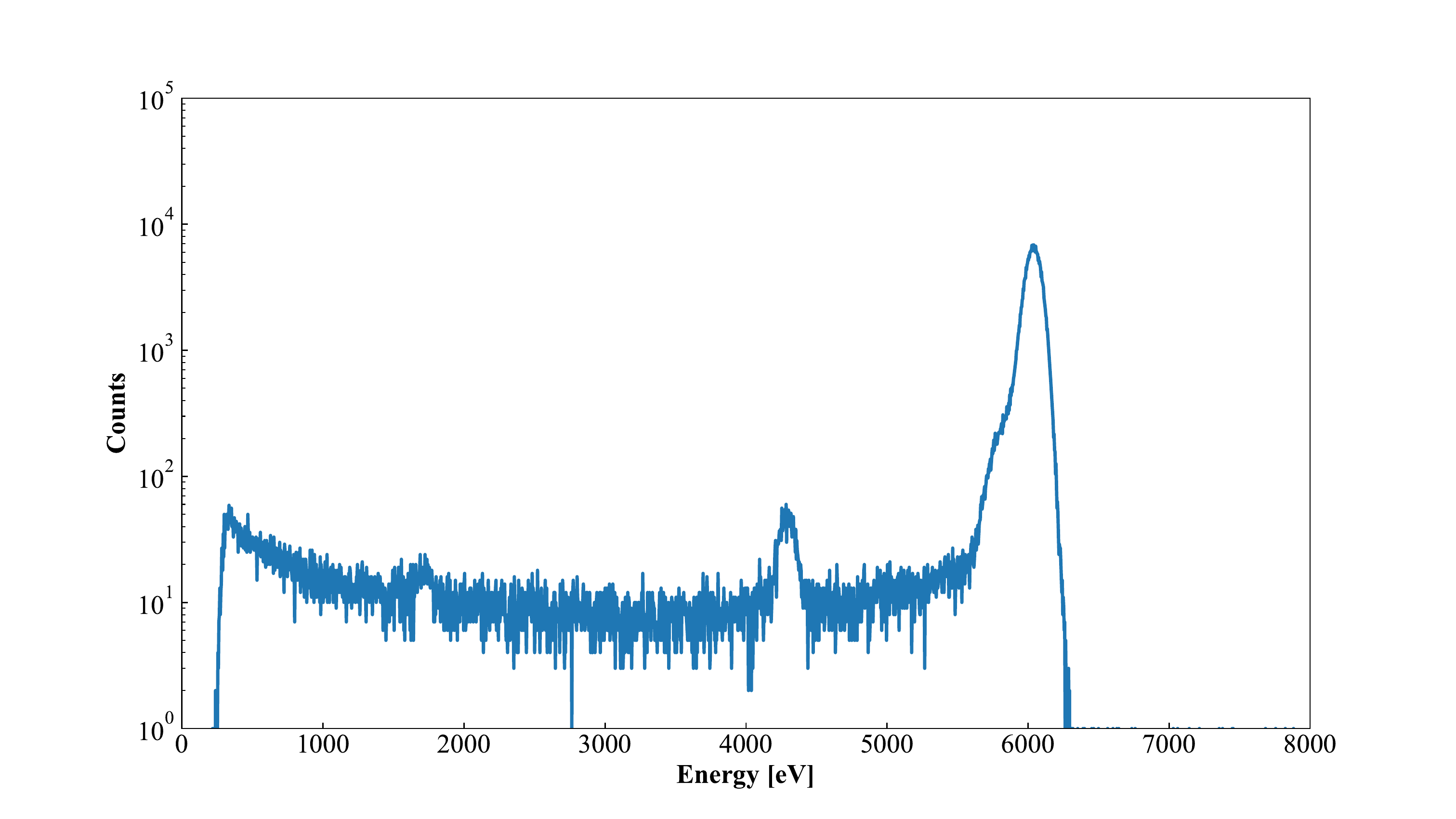}
\caption{Spectrum of the 6 keV X-ray beam produced by the double-crystal monochromator and measured by HXMT-LE. }
\label{figCCD236}
\end{figure}

For the detailed energy response calibration of X-ray detectors, a channel cut double-crystal monochromator is adopted to produce adjustable mono-energy X-rays, as shown in Figure~\ref{fig6}. One of the crystals, like KAP/Si111/Si220/Si440, can be installed and replaced on the monochromator so as to cover 1$\sim$40 keV. As Figure~\ref{figCCD236} shows, this source has been successfully applied in the energy response calibration of the CCD236 detector of the Low Energy Telescope onboard HXMT\cite{RSP}. 

\begin{figure}[!htp]
\centering
\includegraphics[width=0.5\textwidth]{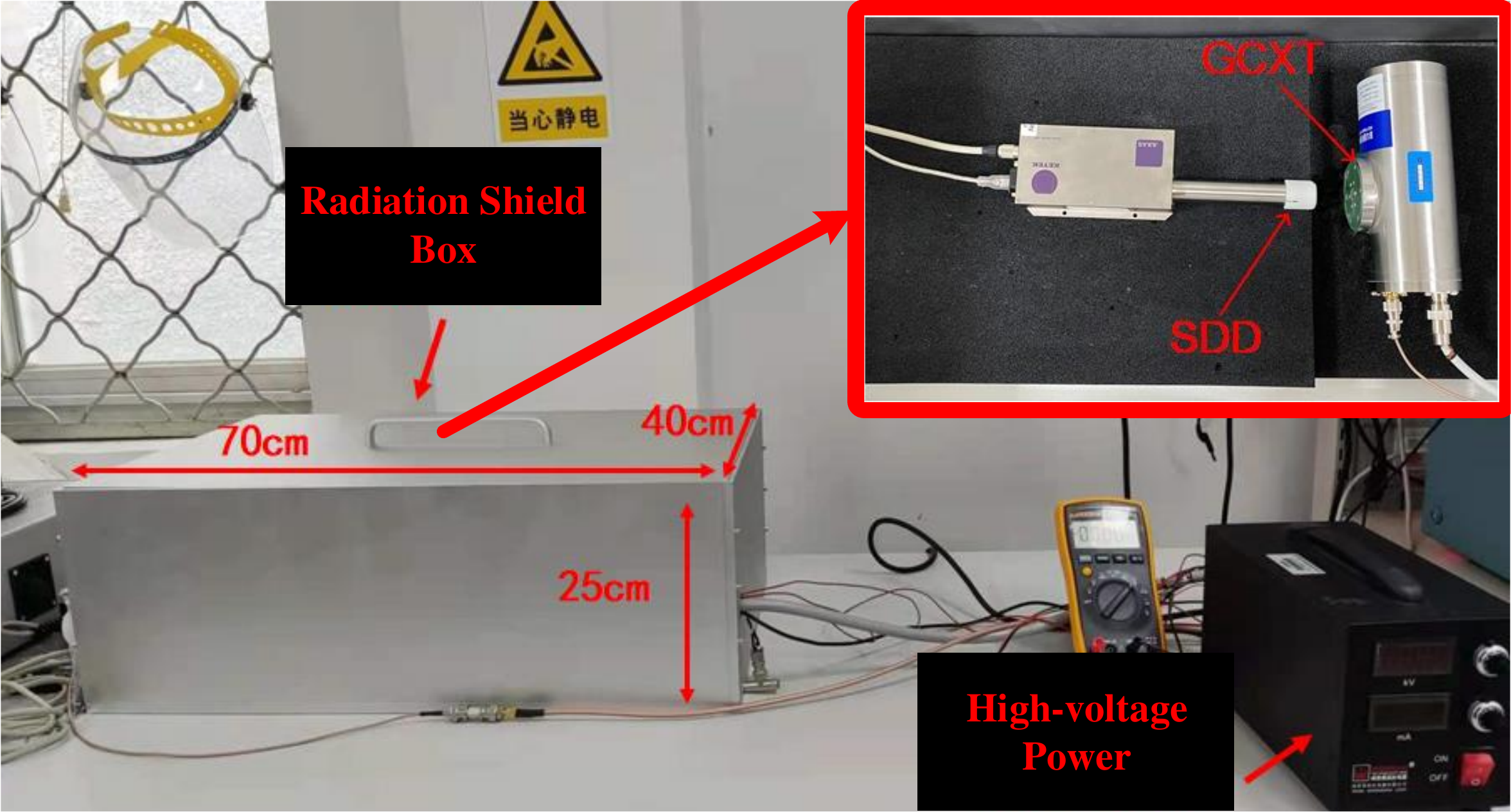}
\caption{The modulated X-ray source in test. The source is designed based on the Grid Controlled X-ray Tube (GCXT) in the radiation shield box.}
\label{figGRB}
\end{figure}

\begin{figure}[!htp]
\centering
\includegraphics[width=0.6\textwidth]{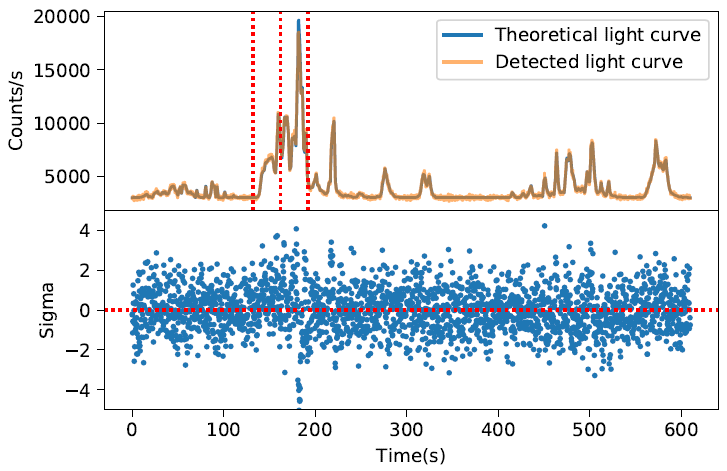}
\caption{Comparison between the theoretical input and detected light curves.}
\label{fig7}
\end{figure}

To calibrate the timing performance of the detector and telescope system, a modulated X-ray source was developed\cite{gammaray} and verified as shown in Figure~\ref{figGRB}. Its time resolution is about 0.9 $\upmu {\rm s}$ with a time delay of about 1.4 $\upmu {\rm s}$. As shown in Figure~\ref{fig7}, the deviation of the detected light curve from the input one is less than $4\upsigma$ at anytime, and the deviation distribution is rather uniform. This modulated source has been applied to calibrate the time performance of GECAM, especially to test the triggering by a gamma ray burst.

\begin{figure}[!htp]
\centering
\includegraphics[width=0.3\textwidth]{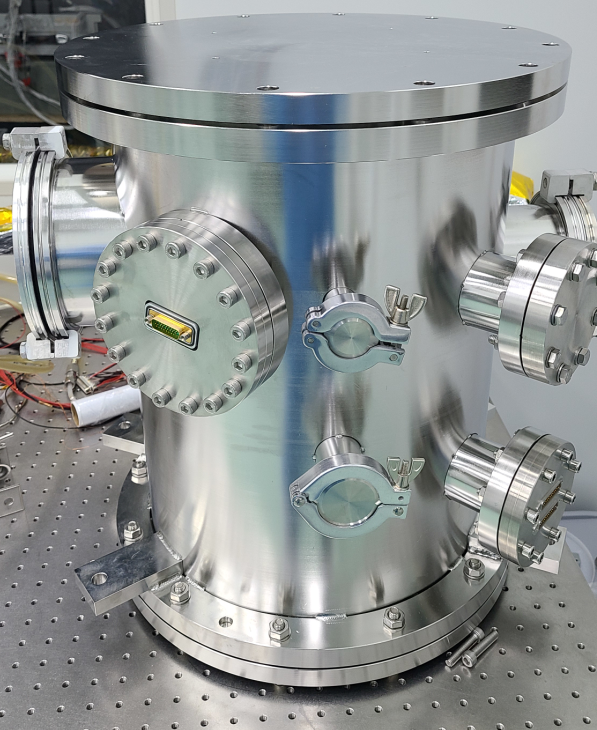}
\caption{The polarized X-ray source to be connected to the 100XF. }
\label{fig8}
\end{figure}

In order to calibrate the polarimetry performance of X-ray telescopes and focal plane detectors, a polarized X-ray source has been developed. As shown in Figure~\ref{fig8}, a small vacuum chamber accommodates several crystals and the corresponding micro-focus X-ray tubes to achieve highly polarized X-ray beam at 2.7, 4.5 and 6.4 keV, with several at other energies under development. We note here that the polarized source is the key device in the polarized performance calibration of Polarimetry Focusing Array (PFA) on board eXTP and some small polarimetry cube satellites. 

\subsection{X-ray detectors}\label{subsec2-4}

There are several standard detectors (cameras) inside the large instrument chamber to measure the spectral, imaging and timing properties of the instruments in test, including an Andor DX436 camera, an Amptek X123 SDDs (C1, C2, Be windows), a Color X-ray Camera (CXC) and a big sCMOS camera. 

\begin{figure}[!htp]
\centering
\includegraphics[width=0.4\textwidth]{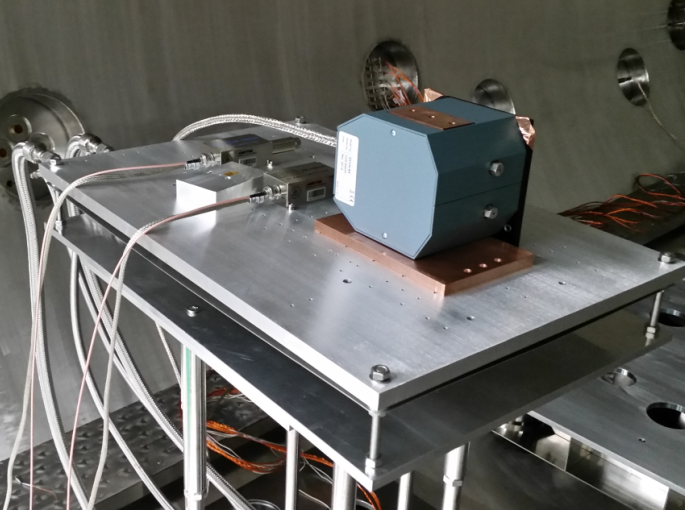}
\caption{The Andor DX436 camera on the 3-D movable stage. The plate under the camera is for cooling. }
\label{fig9}
\end{figure}

The DX436 is a typical soft X-ray sensitive CCD camera  with an area of 27.6 mm$\times$27.6 mm and 2048$\times$2048 pixels (13.5 $\upmu$m for each one) and its frame readout time for the whole image is 1 second. As shown in Figure~\ref{fig9}, the camera is cooled by liquid nitrogen, and the thermal control system can keep the camera working under -60\textcelsius\ stably.

\begin{figure}[!htp]
\centering
\includegraphics[width=0.4\textwidth]{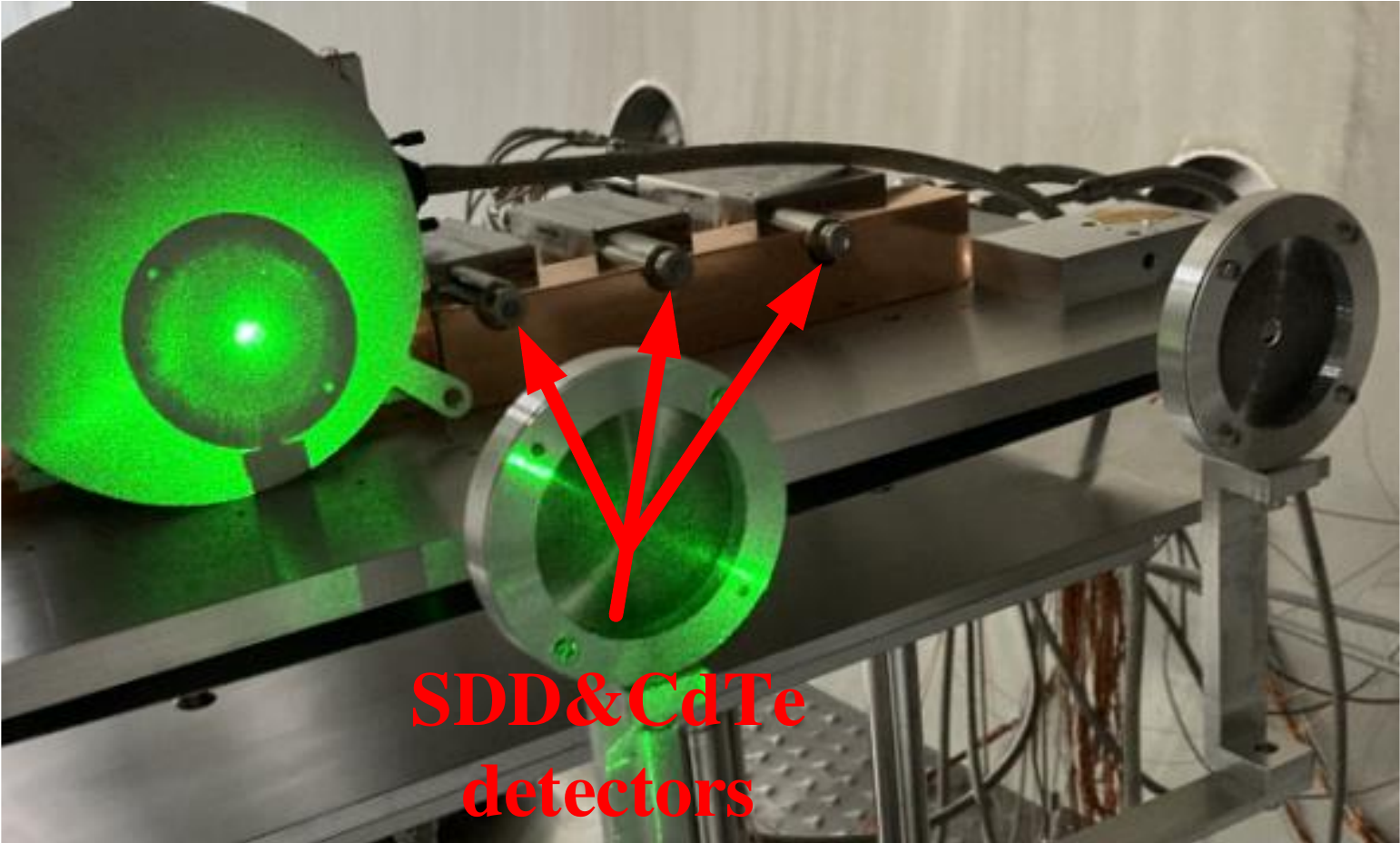}
\caption{Two SDDs and one CdTe are applied to monitor the X-ray beam during the test of a Micro-Slot-Optics telescope. }
\label{fig10}
\end{figure}

As shown in Figure~\ref{fig10}, there are also several SDDs and CdTe detectors, in which one SDD has been calibrated in Physikalisch-Technische Bundesanstalt (PTB), meaning that it can be used to calibrate the absolute Quantum Efficiency (QE) of a telescope or detector. The CdTe detector is used to monitor the high energy X-ray beam (more than 30 keV).

\begin{figure}[!htp]
\centering
\includegraphics[width=0.4\textwidth]{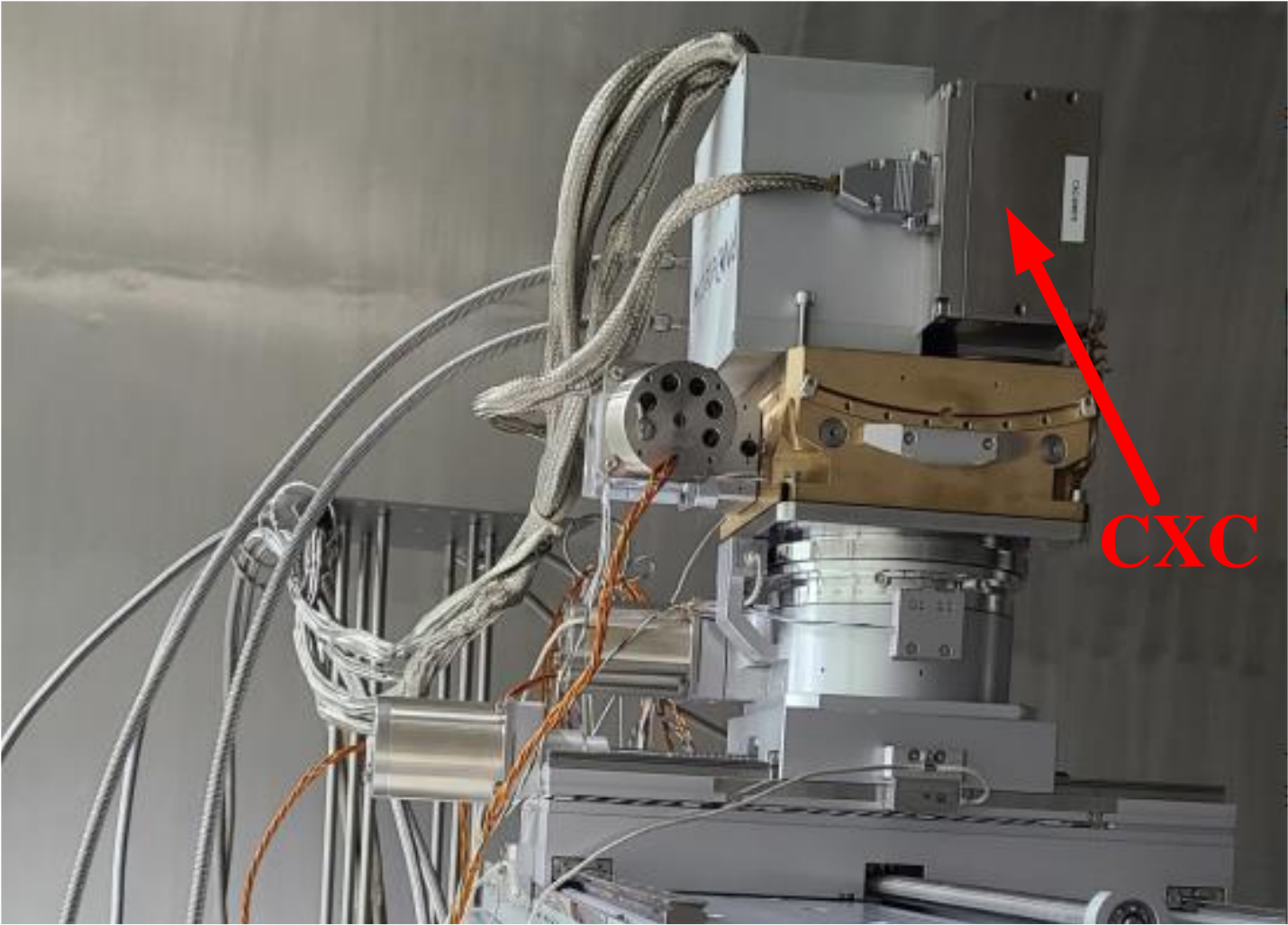}
\caption{The CXC installed on the 6-D movable stage.}
\label{fig11}
\end{figure}

For the point spread function (PSF) measurements and precise measurement of effective area, a Color X-ray Camera (CXC) was developed, based on the pnCCD\cite{PNDetectorCXC}. As shown in Figure~\ref{fig11}, the CXC is configured with complex back-end readout electronics.

\newcommand{\tabincell}[2]{\begin{tabular}{@{}#1@{}}#2\end{tabular}}
\begin{table}[!htp]
\begin{center}
\begin{minipage}{190pt}
\caption{Key performance of the Color X-ray Camera}\label{tab3}%
\begin{tabular}{ll}
\toprule
Name & Value\\
\midrule
Energy range    & 0.2$\sim$30 keV\\
Energy resolution    & $\sim$145 eV @ 6.0 keV\\
Readout time    & \tabincell{l}{1000 fps for full frame\\3400 fps for folded readout}\\
QE    &\tabincell{l}{ 80\% for 500 eV\\ 90\% for 10 keV\\ 30\% for 20 keV}\\
CTE    & \textgreater0.999999\\
\botrule
\end{tabular}
\end{minipage}
\end{center}
\end{table}

Benefited from many advantages of the CXC, such as broad sensitive energy band, superior energy resolution, and the extraordinary high QE and Charge Transfer Efficiency (CTE) (shown in Table~\ref{tab3}), it is perfectly suitable for calibration of optics like Wolter-I and MPO.

\begin{figure}[!htp]
\centering
\includegraphics[width=0.5\textwidth]{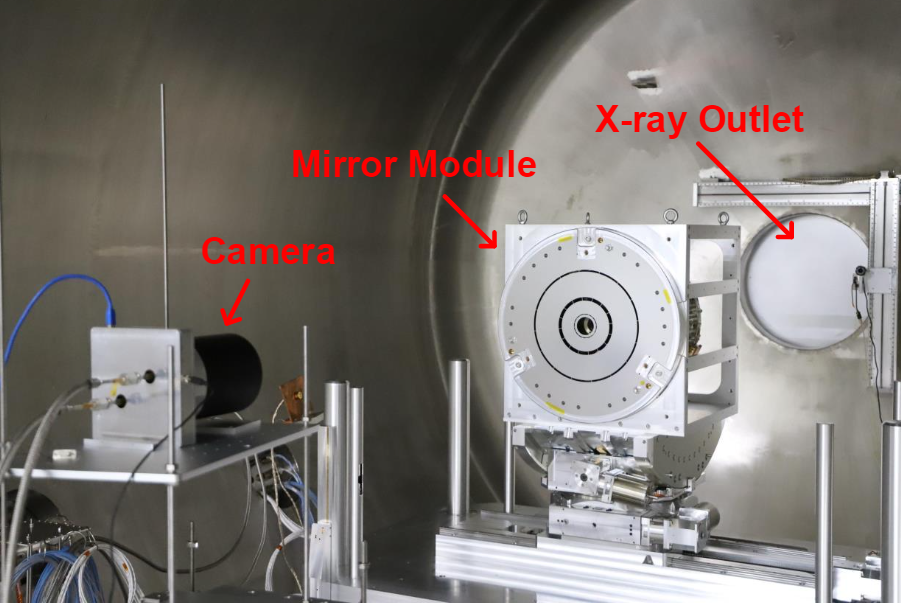}
\caption{The big sCMOS X-ray camera in the EP-FXT focusing mirror module test. }
\label{fig12}
\end{figure}

However, for X-ray optics under development, there are some unknown distortions that make their performance not very good or even difficult to understand. A camera with large detection area that can acquire the whole X-ray reflection image will help to diagnose the defects. Therefore, a large format sCMOS camera has been developed and verified, as shown in Figure~\ref{fig12}\cite{cmos}. Its size is 61 mm$\times$61 mm, which is large enough to get the single reflection of Wolter-I mirror and the whole image of focusing MPO. 

\subsection{Contamination control}\label{subsec2-4}

The X-ray optics, payloads and satellites tested in the 100XF are extremely sensitive to molecular and particle contaminants. Hence the contamination control is a very important and tough work for the whole facility and during all the tests. For example, molecular and particle contaminants can seriously degrade the HEW (Half Energy Width) of the X-ray optics through X-ray scattering with particles, and reduce the effective area by absorption\cite{contamination}. The configuration of the pump sets of 100XF, such as dry pumps, molecular pumps and cryogenic pumps are therefore oil-free. The movable stages, cables, cooling water tubes can all meet the contamination-control requirements too.

As shown in Figure~\ref{fig2}(b), a class-10000 (ISO6) and a class-1000 (ISO7) cleanroom and two dressing rooms are also built for the 100XF. Before enter the chamber operation room, operators need to pass the clean rooms level by level and change the corresponding level of clean clothes as illustrated by the green dotted line in Figure~\ref{fig2}(b). In order to ensure the cleanliness of the tested telescope, especially the optical system, operations of all optical elements are strictly limited to be in the class-1000 (ISO6) cleanroom. On the contrary, the operation of test instruments that do not require high cleanliness, such as detectors, can be completed in the class-10000 (ISO7) environment.

Inside the instrument chamber the particle contamination is monitored by PFO sample (ESA standard) and the molecular contamination is measured by quartz crystal microbalance all the time.

\section{Test and simulation results}\label{sec3}

\begin{figure}[!htp]
\centering
\includegraphics[width=0.5\textwidth]{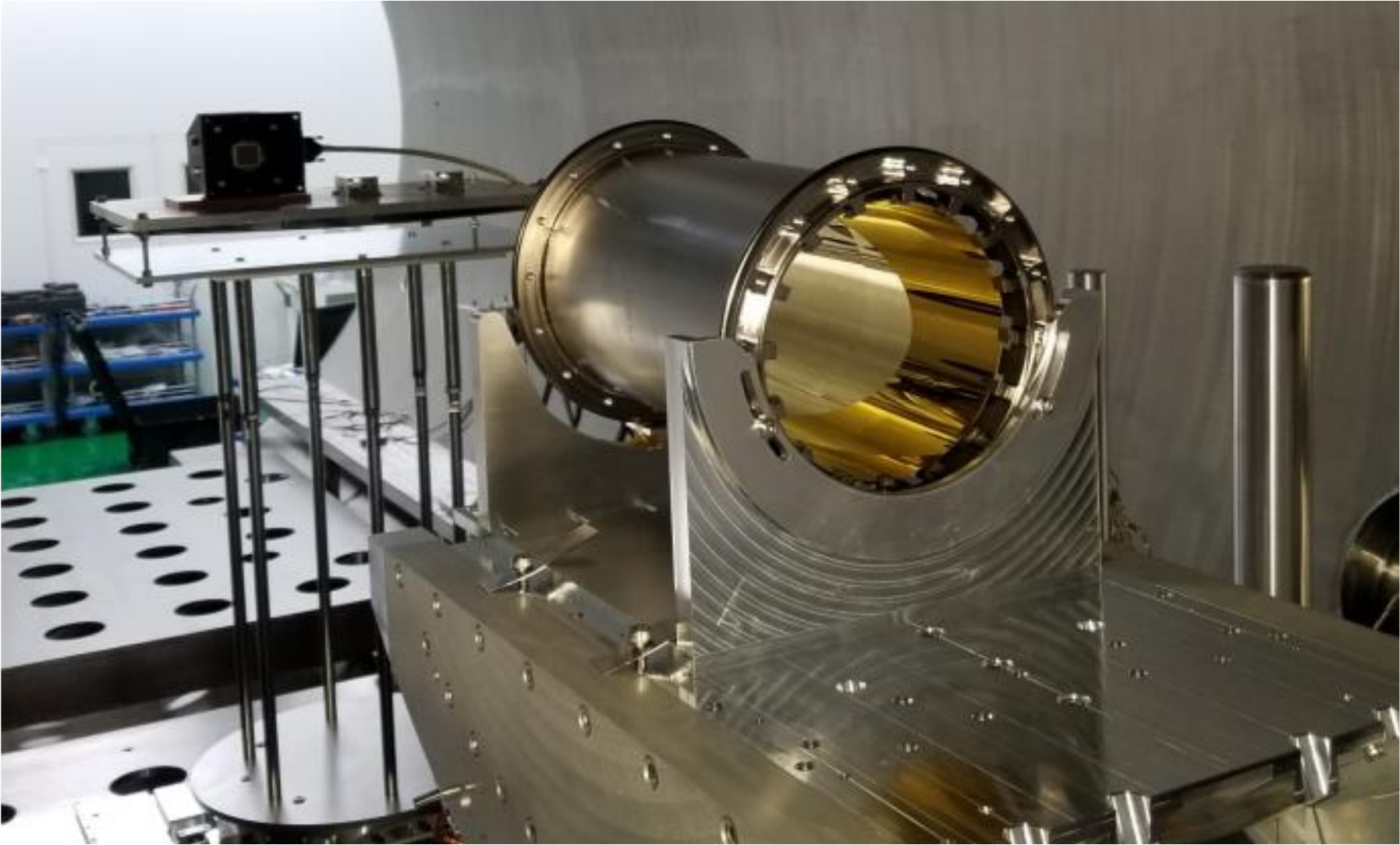}
\caption{Test of a standard Wolter-I shell in the large instrument chamber.}
\label{fig14}
\end{figure}

So far, several EP-FXT mirror modules and eXTP-SFA mirror shells have been tested in the 100XF. Some small optics like MSO, MPO were also tested. Figure~\ref{fig14} shows the test of a standard Wolter-I mirror shell, the results of which are almost identical to that obtained at the PANTER facility in Munich\cite{STMPanterTest}.

\begin{figure}[!htp]
\centering
\includegraphics[width=0.5\textwidth]{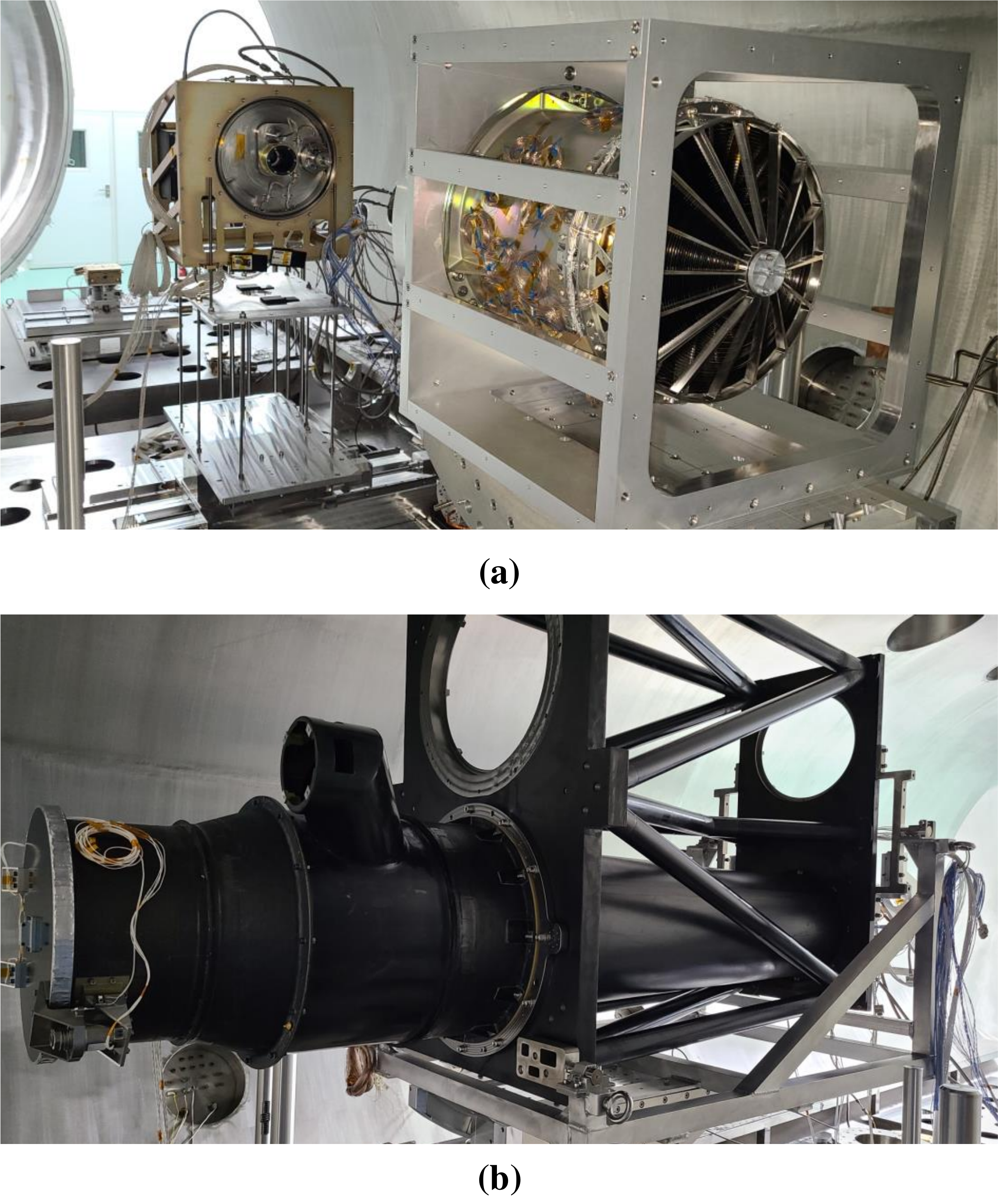}
\caption{(a) Separated test of focusing mirror and focal plane detector pnCCD of FXT. (b) Joint test of the FXT (only one telescope is tested). The mirror STM provided by ESA is installed in the facility. STM contains all 54 shells of FXT, but only 6 shells (\#1,\#24$\sim$27 and \#54) are reflective of X-ray.}
\label{fig15}
\end{figure}

Mirror modules of EP-FXT\cite{FXTchen} have been calibrated in detail in the facility. As shown in Figure~\ref{fig15}, the FXT mirror and focal plane detector are calibrated in the chamber separately and jointly. In the joint test, only one of the two FXT telescopes was installed on the supporting structure as Figure~\ref{fig15}(b) shows.

\begin{figure}[!htp]
\centering
\includegraphics[width=0.8\textwidth]{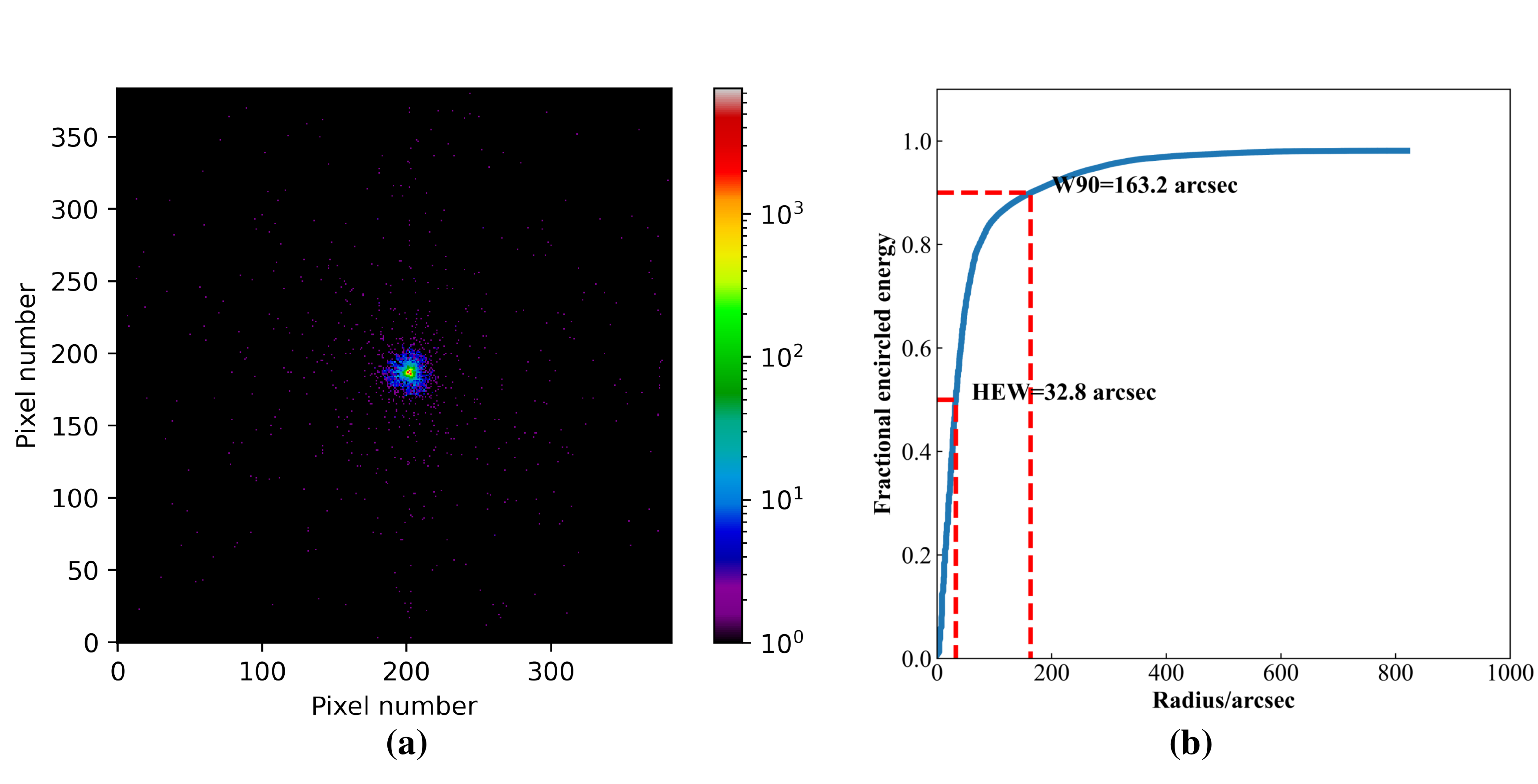}
\caption{(a) Focal plane spot of the FXT separated test. The triangle spot is caused by some irreversible deformation which is also mentioned in the MPE test report. (b) The EEF of the test spot @ 1.49 keV.}
\label{stm}
\end{figure}

A series of basic items such as angular resolution and effective area of the structural thermal model (STM) of FXT are calibrated. Figure~\ref{stm}(a) shows a quite sharp spot. According to the Energy Encircled Function (EEF) inferred from the image and shown in Figure~\ref{stm}(b), the angular resolution of the model is about $32.8^{\prime \prime}$ represented by HEW, and width of 90\% energy (W90) is $163.2^{\prime \prime}$. In addition, the effective area of the STM is measured as $41.11\pm0.55\ {\rm cm^{2}}$\ @ 1.49 keV, which is basically the same as PANTER's result\cite{STMPanterTest}.

\begin{figure}[!htp]
\centering
\includegraphics[width=0.6\textwidth]{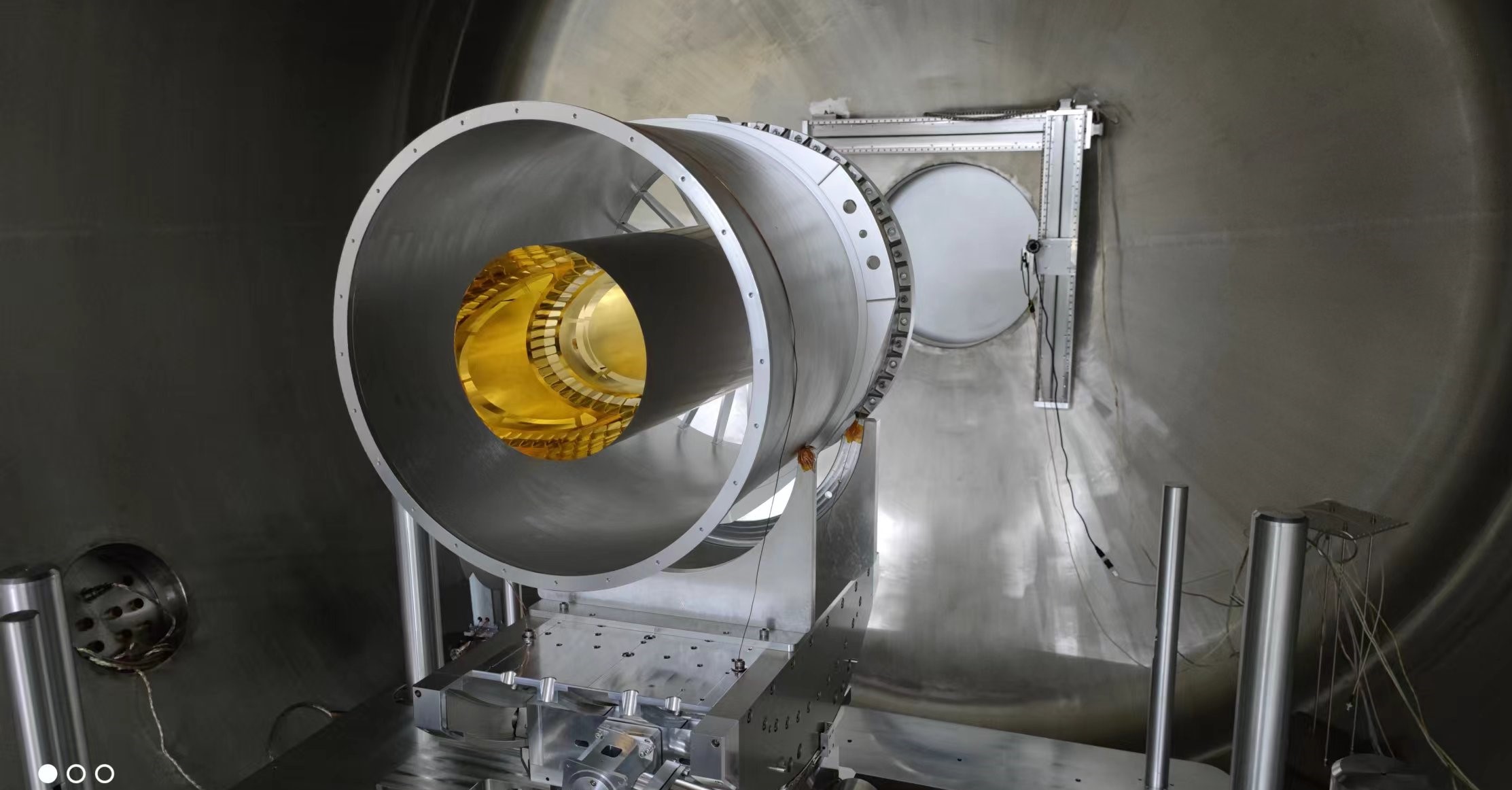}
\caption{Mirror sample of eXTP installed in the chamber of 100XF.}
\label{fig16}
\end{figure}

\begin{figure}[!htp]
\centering
\includegraphics[width=0.7\textwidth]{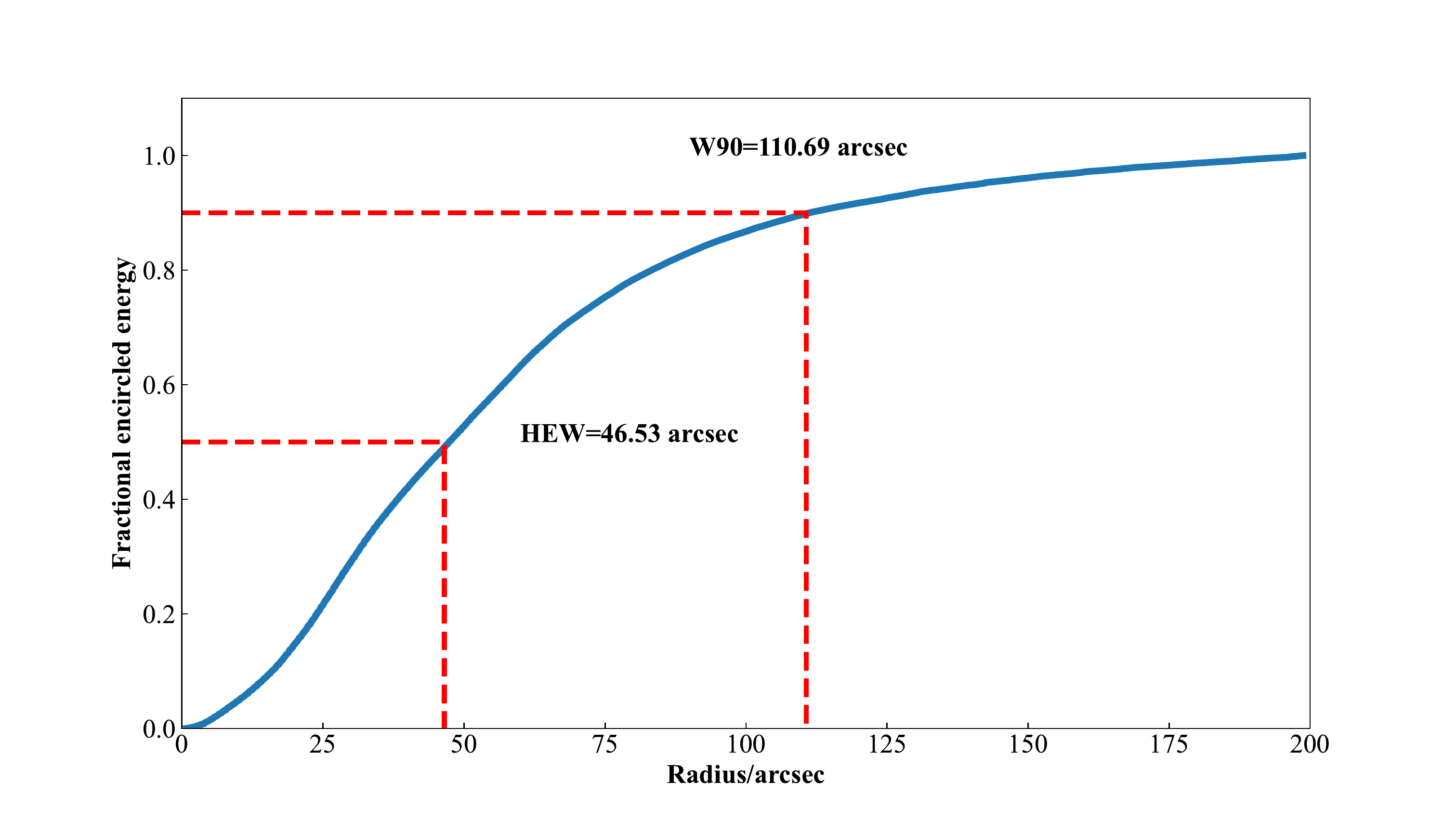}
\caption{The EEF of the test spot  eXTP \#45 shell @ 1.49 keV.}
\label{fig17}
\end{figure}

Tests of the experimental mirror shells for eXTP are also carried out in the 100XF (Figure~\ref{fig16}). As Figure~\ref{fig17} shows, the shell sample has an imaging performance of HEW about $46.53^{\prime \prime}$ and W90 of $110.69^{\prime \prime}$, respectively, which meet the requirements of eXTP-SFA.

Meanwhile, the measurement of the effective area of shell \#45 is carried out with the CXC. The measured area at 4.5 keV is 30\% smaller than that obtained with simulation. However, after the correction of the beam divergence and distance bias between the optics and CXC, the discrepancy becomes less than 10\%,  similar to those of most X-ray mirrors\cite{XMM-Newton} in the past. We speculate that this discrepancy could be due to the imperfection of the surface and contamination.

In summary, the above introduction shows that the 100XF is ready for test and calibration of various X-ray detectors and telescopes. The test results of the X-ray mirrors of EP--FXT and eXTP--SFA also show that the calibrations carried out with the 100XF are precise and reliable. Given the size of the chamber, 100XF can test x-ray mirrors with focal length up to 6 m and diameter smaller than 60 cm. In addition, due to the beam-line finite length and detector pixel size, the best angular resolution of grazing incidence optics achievable in the tests is $5^{\prime \prime}$ HEW.  We expect further improvements on 100XF so that it can satisfy all the calibration requirements of future X-ray missions.
\backmatter

\bmhead{Author’s Contribution}

YW is responsible for the construction of the facility and wrote the main manuscript text. YZ helped to the writing and prepared figure 8, 18 and 20. ZZ prepared the figure 2, 3, 4 and the corresponding description. DH and XZ prepared the figure 5, 6, 14 and the corresponding description. XY prepared the figure 11. CC prepared the figure 9, 10 and the corresponding description. XL and HX reviewed figure 1 and gave suggestions. JM help to check the description of all figures. YC, GW and SZ prepared the figure 7 and the corresponding description. FL and SZ reviewed the language grammar and gave some helpful suggestions. YC reviewed the test results of EP-FXT. YX reviewed the test results of eXTP. All authors read and approved the final manuscript.

\bmhead{Acknowledgments}

We are grateful for the helpful discussion with Vadim Burwitz and Peter Friedrich from Max-Planck-Institut f\"{u}r extraterrestrische Physik.

\section*{Declarations}

The authors declare that they have no conflict of interest.

\bmhead{Funding}

This work is the supported by the National Natural Science Foundation of China (GrantNo. U1838201 and U1838202), Strategic Priority Research Program on Space Science, the Chinese Academy of Sciences (No. XDA1531010301 and XDA15020500), and partially supported by International Partnership Program of Chinese Academy of Sciences (Grant No.113111KYSB20190020).

\bibliography{reference}


\begin{thebibliography}{24}
\ifx \bisbn   \undefined \def \bisbn  #1{ISBN #1}\fi
\ifx \binits  \undefined \def \binits#1{#1}\fi
\ifx \bauthor  \undefined \def \bauthor#1{#1}\fi
\ifx \batitle  \undefined \def \batitle#1{#1}\fi
\ifx \bjtitle  \undefined \def \bjtitle#1{#1}\fi
\ifx \bvolume  \undefined \def \bvolume#1{\textbf{#1}}\fi
\ifx \byear  \undefined \def \byear#1{#1}\fi
\ifx \bissue  \undefined \def \bissue#1{#1}\fi
\ifx \bfpage  \undefined \def \bfpage#1{#1}\fi
\ifx \blpage  \undefined \def \blpage #1{#1}\fi
\ifx \burl  \undefined \def \burl#1{\textsf{#1}}\fi
\ifx \doiurl  \undefined \def \doiurl#1{\url{https://doi.org/#1}}\fi
\ifx \betal  \undefined \def \betal{\textit{et al.}}\fi
\ifx \binstitute  \undefined \def \binstitute#1{#1}\fi
\ifx \binstitutionaled  \undefined \def \binstitutionaled#1{#1}\fi
\ifx \bctitle  \undefined \def \bctitle#1{#1}\fi
\ifx \beditor  \undefined \def \beditor#1{#1}\fi
\ifx \bpublisher  \undefined \def \bpublisher#1{#1}\fi
\ifx \bbtitle  \undefined \def \bbtitle#1{#1}\fi
\ifx \bedition  \undefined \def \bedition#1{#1}\fi
\ifx \bseriesno  \undefined \def \bseriesno#1{#1}\fi
\ifx \blocation  \undefined \def \blocation#1{#1}\fi
\ifx \bsertitle  \undefined \def \bsertitle#1{#1}\fi
\ifx \bsnm \undefined \def \bsnm#1{#1}\fi
\ifx \bsuffix \undefined \def \bsuffix#1{#1}\fi
\ifx \bparticle \undefined \def \bparticle#1{#1}\fi
\ifx \barticle \undefined \def \barticle#1{#1}\fi
\bibcommenthead
\ifx \bconfdate \undefined \def \bconfdate #1{#1}\fi
\ifx \botherref \undefined \def \botherref #1{#1}\fi
\ifx \url \undefined \def \url#1{\textsf{#1}}\fi
\ifx \bchapter \undefined \def \bchapter#1{#1}\fi
\ifx \bbook \undefined \def \bbook#1{#1}\fi
\ifx \bcomment \undefined \def \bcomment#1{#1}\fi
\ifx \oauthor \undefined \def \oauthor#1{#1}\fi
\ifx \citeauthoryear \undefined \def \citeauthoryear#1{#1}\fi
\ifx \endbibitem  \undefined \def \endbibitem {}\fi
\ifx \bconflocation  \undefined \def \bconflocation#1{#1}\fi
\ifx \arxivurl  \undefined \def \arxivurl#1{\textsf{#1}}\fi
\csname PreBibitemsHook\endcsname

%%% 1
\bibitem{2007HXMT}
\begin{barticle}
\bauthor{\bsnm{Li}, \binits{T.P.}}:
\batitle{{HXMT}: A {C}hinese high-energy astrophysics mission}.
\bjtitle{Nuclear Physics B - Proceedings Supplements}
\bvolume{166},
\bfpage{131}--\blpage{139}
(\byear{2007})
\end{barticle}
\endbibitem

%%% 2
\bibitem{2020HXMT}
\begin{botherref}
\oauthor{\bsnm{Zhang}, \binits{S.N.}},
\oauthor{\bsnm{Li}, \binits{T.P.}},
\oauthor{\bsnm{Lu}, \binits{F.J.}},
\oauthor{\bsnm{Song}, \binits{L.M.}},
\oauthor{\bsnm{Zhuang}, \binits{R.L.}}:
Overview to the {H}ard {X}-ray {M}odulation {T}elescope
  (\textit{{I}nsight}-{HXMT}) {S}atellite.
Science China: Physics, Mechanics and Astronomy
\textbf{63}(4)
(2020)
\end{botherref}
\endbibitem

%%% 3
\bibitem{2018EP}
\begin{bchapter}
\bauthor{\bsnm{Yuan}, \binits{W.}},
\bauthor{\bsnm{Chen}, \binits{Z.}},
\bauthor{\bsnm{Ling}, \binits{Z.}},
\bauthor{\bsnm{Zhao}, \binits{D.}},
\bauthor{\bsnm{Wei}, \binits{C.}}:
\bctitle{Einstein {P}robe: a lobster-eye telescope for monitoring the x-ray
  sky}.
In: \bbtitle{Space Telescopes and Instrumentation 2018: Ultraviolet to Gamma
  Ray}
(\byear{2018})
\end{bchapter}
\endbibitem

%%% 4
\bibitem{GECAM}
\begin{botherref}
\oauthor{\bsnm{Xiong}, \binits{S.}}:
{GECAM} gamma-ray all-sky monitor.
SCIENTIA SINICA Physica, Mechanica \& Astronomica
\textbf{50}(12)
(2020)
\end{botherref}
\endbibitem

%%% 5
\bibitem{2019eXTP}
\begin{barticle}
\bauthor{\bsnm{Zhang}, \binits{S.N.}},
\bauthor{\bsnm{Hernanz}, \binits{M.}},
\bauthor{\bsnm{Santangelo}, \binits{A.}},
\bauthor{\bsnm{Feroci}, \binits{M.}},
\bauthor{\bsnm{Xu}, \binits{Y.P.}},
\bauthor{\bsnm{Lu}, \binits{F.J.}},
\bauthor{\bsnm{Chen}, \binits{Y.}},
\bauthor{\bsnm{Feng}, \binits{H.}},
\bauthor{\bsnm{Zhang}, \binits{S.}},
\bauthor{\bsnm{Brandt}, \binits{S.}}:
\batitle{The enhanced {X}-ray {T}iming and {P}olarimetry mission—e{XTP}}.
\bjtitle{SCIENTIA SINICA Physica, Mechanica \& Astronomica}
\bvolume{62}(\bissue{2}),
\bfpage{25}
(\byear{2019})
\end{barticle}
\endbibitem

%%% 6
\bibitem{panter1979}
\begin{barticle}
\bauthor{\bsnm{Bernd}, \binits{A.}},
\bauthor{\bsnm{Heinrich}, \binits{B.}},
\bauthor{\bsnm{Karl-Heinz}, \binits{S.}},
\bauthor{\bsnm{Joachim}, \binits{T.}}:
\batitle{{X-ray test facilities at Max-Planck-Institut Garching}}.
\bjtitle{SPIE}
\bvolume{184},
\bfpage{234}--\blpage{238}
(\byear{1979})
\end{barticle}
\endbibitem

%%% 7
\bibitem{panter}
\begin{barticle}
\bauthor{\bsnm{Freyberg}, \binits{M.J.}},
\bauthor{\bsnm{Bräuninger}, \binits{H.}},
\bauthor{\bsnm{Burkert}, \binits{W.}},
\bauthor{\bsnm{Hartner}, \binits{G.D.}},
\bauthor{\bsnm{Citterio}, \binits{O.}},
\bauthor{\bsnm{Mazzoleni}, \binits{F.}},
\bauthor{\bsnm{Pareschi}, \binits{G.}},
\bauthor{\bsnm{Spiga}, \binits{D.}},
\bauthor{\bsnm{Romaine}, \binits{S.}},
\bauthor{\bsnm{Gorenstein}, \binits{P.}},
\bauthor{\bsnm{Ramsey}, \binits{B.D.}}:
\batitle{{The MPE X-ray test facility PANTER: Calibration of hard X-ray (15--50
  kev) optics}}.
\bjtitle{Exp Astron}
\bvolume{20},
\bfpage{405}--\blpage{412}
(\byear{2005})
\end{barticle}
\endbibitem

%%% 8
\bibitem{XMM}
\begin{barticle}
\bauthor{\bsnm{{Jansen, F.}}},
\bauthor{\bsnm{{Lumb, D.}}},
\bauthor{\bsnm{{Altieri, B.}}},
\bauthor{\bsnm{{Clavel, J.}}},
\bauthor{\bsnm{{Ehle, M.}}},
\bauthor{\bsnm{{Erd, C.}}},
\bauthor{\bsnm{{Gabriel, C.}}},
\bauthor{\bsnm{{Guainazzi, M.}}},
\bauthor{\bsnm{{Gondoin, P.}}},
\bauthor{\bsnm{{Much, R.}}},
\bauthor{\bsnm{{Munoz, R.}}},
\bauthor{\bsnm{{Santos, M.}}},
\bauthor{\bsnm{{Schartel, N.}}},
\bauthor{\bsnm{{Texier, D.}}},
\bauthor{\bsnm{{Vacanti, G.}}}:
\batitle{{XMM-Newton observatory - I. The spacecraft and operations}}.
\bjtitle{A\&A}
\bvolume{365}(\bissue{1}),
\bfpage{1}--\blpage{6}
(\byear{2001}).
\doiurl{10.1051/0004-6361:20000036}
\end{barticle}
\endbibitem

%%% 9
\bibitem{eROSITA}
\begin{barticle}
\bauthor{\bsnm{Predehl}, \binits{P.}},
\bauthor{\bsnm{Böhringer}, \binits{H.}},
\bauthor{\bsnm{Brunner}, \binits{H.}},
\bauthor{\bsnm{Brusa}, \binits{M.}},
\bauthor{\bsnm{Burwitz}, \binits{V.}},
\bauthor{\bsnm{Cappelluti}, \binits{N.}},
\bauthor{\bsnm{Churazov}, \binits{E.}},
\bauthor{\bsnm{Dennerl}, \binits{K.}},
\bauthor{\bsnm{Freyberg}, \binits{M.}},
\bauthor{\bsnm{Friedrich}, \binits{P.}},
\bauthor{\bsnm{Hasinger}, \binits{G.}},
\bauthor{\bsnm{Kendziorra}, \binits{E.}},
\bauthor{\bsnm{Kreykenbohm}, \binits{I.}},
\bauthor{\bsnm{Schmid}, \binits{C.}},
\bauthor{\bsnm{Wilms}, \binits{J.}},
\bauthor{\bsnm{Lamer}, \binits{G.}},
\bauthor{\bsnm{Meidinger}, \binits{N.}},
\bauthor{\bsnm{Mühlegger}, \binits{M.}},
\bauthor{\bsnm{Pavlinsky}, \binits{M.}},
\bauthor{\bsnm{Robrade}, \binits{J.}},
\bauthor{\bsnm{Santangelo}, \binits{A.}},
\bauthor{\bsnm{Schmitt}, \binits{J.}},
\bauthor{\bsnm{Schwope}, \binits{A.}},
\bauthor{\bsnm{Steinmetz}, \binits{M.}},
\bauthor{\bsnm{Strüder}, \binits{L.}},
\bauthor{\bsnm{Sunyaev}, \binits{R.}},
\bauthor{\bsnm{Tenzer}, \binits{C.}}:
\batitle{erosita on srg}.
\bjtitle{AIP Conference Proceedings}
\bvolume{1248}(\bissue{1}),
\bfpage{543}--\blpage{548}
(\byear{2010})
{\href{https://arxiv.org/abs/https://aip.scitation.org/doi/pdf/10.1063/1.3475336}{{https://aip.scitation.org/doi/pdf/10.1063/1.3475336}}}.
\doiurl{10.1063/1.3475336}
\end{barticle}
\endbibitem

%%% 10
\bibitem{XRCF}
\begin{barticle}
\bauthor{\bsnm{Jeff}, \binits{K.}},
\bauthor{\bsnm{Mark}, \binits{B.}},
\bauthor{\bsnm{Jay}, \binits{C.}},
\bauthor{\bsnm{Ron}, \binits{E.}},
\bauthor{\bsnm{Harlan}, \binits{H.}},
\bauthor{\bsnm{William}, \binits{H.}},
\bauthor{\bsnm{Jeff}, \binits{M.}},
\bauthor{\bsnm{Kevin}, \binits{R.}},
\bauthor{\bsnm{Richar}, \binits{S.}},
\bauthor{\bsnm{Ernie}, \binits{W.}}:
\batitle{{Improved Cryogenic Testing Capability at Marshall Space Flight
  Center's X-ray Cryogenic Facility}}.
\bjtitle{SPIE}
\bvolume{6265},
\bfpage{62654}--\blpage{1626548}
(\byear{2006})
\end{barticle}
\endbibitem

%%% 11
\bibitem{Chandra}
\begin{botherref}
\oauthor{\bsnm{Weisskopf}, \binits{M.C.}},
\oauthor{\bsnm{Tananbaum}, \binits{H.D.}},
\oauthor{\bsnm{Speybroeck}, \binits{L.P.V.}},
\oauthor{\bsnm{O'Dell}, \binits{S.L.}}:
{Chandra X-ray Observatory (CXO): overview}
\textbf{4012},
2--16
(2000).
\doiurl{10.1117/12.391545}.
International Society for Optics and Photonics
\end{botherref}
\endbibitem

%%% 12
\bibitem{XACT}
\begin{botherref}
\oauthor{\bsnm{Barbera}, \binits{M.}},
\oauthor{\bsnm{Candia}, \binits{R.}},
\oauthor{\bsnm{Collura}, \binits{A.}},
\oauthor{\bsnm{Cicca}, \binits{G.D.}},
\oauthor{\bsnm{Pelliciari}, \binits{C.}},
\oauthor{\bsnm{Sciortino}, \binits{S.}},
\oauthor{\bsnm{Varisco}, \binits{S.}}:
{The Palermo XACT facility: a new 35 m long soft x-ray beam-line for
  development and calibration of next-generation x-ray observatories}
\textbf{6266},
62663
(2006).
\doiurl{10.1117/12.673004}.
International Society for Optics and Photonics
\end{botherref}
\endbibitem

%%% 13
\bibitem{BEaTriX}
\begin{botherref}
\oauthor{\bsnm{Pelliciari}, \binits{C.}},
\oauthor{\bsnm{Spiga}, \binits{D.}},
\oauthor{\bsnm{Bonnini}, \binits{E.}},
\oauthor{\bsnm{Buffagni}, \binits{E.}},
\oauthor{\bsnm{Ferrari}, \binits{C.}},
\oauthor{\bsnm{Pareschi}, \binits{G.}},
\oauthor{\bsnm{Tagliaferri}, \binits{G.}}:
{BEaTriX, expanded x-ray beam facility for testing modular elements of
  telescope optics: an update}
\textbf{9603},
96031
(2015).
\doiurl{10.1117/12.2188607}.
International Society for Optics and Photonics
\end{botherref}
\endbibitem

%%% 14
\bibitem{vert-x}
\begin{botherref}
\oauthor{\bsnm{Pareschi}, \binits{G.}},
\oauthor{\bsnm{Moretti}, \binits{A.}},
\oauthor{\bsnm{Salmaso}, \binits{B.}},
\oauthor{\bsnm{Sironi}, \binits{G.}},
\oauthor{\bsnm{Tagliaferri}, \binits{G.}},
\oauthor{\bsnm{Uslenghi}, \binits{M.}},
\oauthor{\bsnm{Fiorini}, \binits{M.}},
\oauthor{\bsnm{Attin{\`a}}, \binits{P.}},
\oauthor{\bsnm{Bressan}, \binits{R.}},
\oauthor{\bsnm{Marchiori}, \binits{G.}},
\oauthor{\bsnm{Tordi}, \binits{M.}},
\oauthor{\bsnm{Marioni}, \binits{F.}},
\oauthor{\bsnm{Valsecchi}, \binits{G.}},
\oauthor{\bsnm{Zocchi}, \binits{F.}}:
{A vertical facility based on raster scan configuration for the x-ray
  scientific calibrations of the ATHENA optics}
\textbf{11180},
1118025
(2019).
\doiurl{10.1117/12.2535996}.
International Society for Optics and Photonics
\end{botherref}
\endbibitem

%%% 15
\bibitem{FXT-zhao}
\begin{barticle}
\bauthor{\bsnm{Zhao}, \binits{Z.}},
\bauthor{\bsnm{Wang}, \binits{Y.}},
\bauthor{\bsnm{Zhang}, \binits{L.}},
\bauthor{\bsnm{Chen}, \binits{C.}},
\bauthor{\bsnm{Ma}, \binits{J.}}:
\batitle{{X-ray optical Experiments and Simulation of Wolter-I Focusing
  Mirror}}.
\bjtitle{Optics and Precision Engineering}
\bvolume{27}(\bissue{11}),
\bfpage{2331}--\blpage{2336}
(\byear{2019})
\end{barticle}
\endbibitem

%%% 16
\bibitem{pnccd}
\begin{barticle}
\bauthor{\bsnm{Meidinger}, \binits{N.}},
\bauthor{\bsnm{Andritschke}, \binits{R.}},
\bauthor{\bsnm{Ebermayer}, \binits{S.}},
\bauthor{\bsnm{Elbs}, \binits{J.}},
\bauthor{\bsnm{Hälker}, \binits{O.}},
\bauthor{\bsnm{Hartmann}, \binits{R.}},
\bauthor{\bsnm{Herrmann}, \binits{S.}},
\bauthor{\bsnm{Kimmel}, \binits{N.}},
\bauthor{\bsnm{Predehl}, \binits{P.}},
\bauthor{\bsnm{Schächner}, \binits{G.}},
\bauthor{\bsnm{Soltau}, \binits{H.}},
\bauthor{\bsnm{Strüder}, \binits{L.}},
\bauthor{\bsnm{Tiedemann}, \binits{L.}}:
\batitle{{CCD detectors for spectroscopy and imaging of x-rays with the eROSITA
  space telescope}}.
\bjtitle{Proc of SPIE}
\bvolume{7435},
\bfpage{9}--\blpage{16}
(\byear{2009})
\end{barticle}
\endbibitem

%%% 17
\bibitem{RSP}
\begin{barticle}
\bauthor{\bsnm{Zhu}, \binits{Y.X.}},
\bauthor{\bsnm{Lu}, \binits{J.B.}},
\bauthor{\bsnm{Li}, \binits{X.B.}},
\bauthor{\bsnm{Liu}, \binits{X.Y.}},
\bauthor{\bsnm{Cui}, \binits{W.W.}},
\bauthor{\bsnm{Han}, \binits{D.W.}},
\bauthor{\bsnm{Wang}, \binits{J.}},
\bauthor{\bsnm{Wang}, \binits{Y.S.}},
\bauthor{\bsnm{Zhao}, \binits{X.F.}},
\bauthor{\bsnm{Zhang}, \binits{S.}},
\bauthor{\bsnm{Wang}, \binits{G.F.}},
\bauthor{\bsnm{Chen}, \binits{Y.P.}},
\bauthor{\bsnm{Yu}, \binits{N.}},
\bauthor{\bsnm{Chen}, \binits{Y.}}:
\batitle{Calibration of the energy response matrix for x-ray detector
  {CCD}236}.
\bjtitle{Journal of Instrumentation}
\bvolume{16}(\bissue{05}),
\bfpage{05016}
(\byear{2021}).
\doiurl{10.1088/1748-0221/16/05/p05016}
\end{barticle}
\endbibitem

%%% 18
\bibitem{gammaray}
\begin{barticle}
\bauthor{\bsnm{Chen}, \binits{C.}},
\bauthor{\bsnm{Xiao}, \binits{S.}},
\bauthor{\bsnm{Xiong}, \binits{S.L.}},
\bauthor{\bsnm{Yu}, \binits{N.}},
\bauthor{\bsnm{Wen}, \binits{X.Y.}},
\bauthor{\bsnm{Gong}, \binits{K.}},
\bauthor{\bsnm{Li}, \binits{X.Q.}},
\bauthor{\bsnm{Li}, \binits{C.Y.}},
\bauthor{\bsnm{Hou}, \binits{D.J.}},
\bauthor{\bsnm{Yang}, \binits{X.T.}},
\bauthor{\bsnm{Zhao}, \binits{Z.J.}},
\bauthor{\bsnm{Zhu}, \binits{Y.X.}},
\bauthor{\bsnm{Zhang}, \binits{D.L.}},
\bauthor{\bsnm{An}, \binits{Z.H.}},
\bauthor{\bsnm{Zhao}, \binits{X.Y.}},
\bauthor{\bsnm{Xu}, \binits{Y.P.}},
\bauthor{\bsnm{Wang}, \binits{Y.S.}}:
\batitle{Design and test of a portable {G}amma-{R}ay {B}urst simulator fo
  {GECAM}}.
\bjtitle{Exp. Astron.}
\bvolume{52},
\bfpage{45}--\blpage{58}
(\byear{2021})
\end{barticle}
\endbibitem

%%% 19
\bibitem{PNDetectorCXC}
\begin{barticle}
\bauthor{\bsnm{Ordavo}, \binits{I.}},
\bauthor{\bsnm{Ihle}, \binits{S.}},
\bauthor{\bsnm{Arkadiev}, \binits{V.}},
\bauthor{\bsnm{Scharf}, \binits{O.}},
\bauthor{\bsnm{Soltau}, \binits{H.}},
\bauthor{\bsnm{Bjeoumikhov}, \binits{A.}},
\bauthor{\bsnm{Bjeoumikhova}, \binits{S.}},
\bauthor{\bsnm{Buzanich}, \binits{G.}},
\bauthor{\bsnm{Gubzhokov}, \binits{R.}},
\bauthor{\bsnm{Günther}, \binits{A.}},
\bauthor{\bsnm{Hartmann}, \binits{R.}},
\bauthor{\bsnm{Holl}, \binits{P.}},
\bauthor{\bsnm{Kimmel}, \binits{N.}},
\bauthor{\bsnm{Kühbacher}, \binits{M.}},
\bauthor{\bsnm{Lang}, \binits{M.}},
\bauthor{\bsnm{Langhoff}, \binits{N.}},
\bauthor{\bsnm{Liebel}, \binits{A.}},
\bauthor{\bsnm{Radtke}, \binits{M.}},
\bauthor{\bsnm{Reinholz}, \binits{U.}},
\bauthor{\bsnm{Riesemeier}, \binits{H.}},
\bauthor{\bsnm{Schaller}, \binits{G.}},
\bauthor{\bsnm{Schopper}, \binits{F.}},
\bauthor{\bsnm{Strüder}, \binits{L.}},
\bauthor{\bsnm{Thamm}, \binits{C.}},
\bauthor{\bsnm{Wedell}, \binits{R.}}:
\batitle{A new pnccd-based color x-ray camera for fast spatial and
  energy-resolved measurements}.
\bjtitle{Nuclear Instruments and Methods in Physics Research Section A:
  Accelerators, Spectrometers, Detectors and Associated Equipment}
\bvolume{654}(\bissue{1}),
\bfpage{250}--\blpage{257}
(\byear{2011}).
\doiurl{10.1016/j.nima.2011.05.080}
\end{barticle}
\endbibitem

%%% 20
\bibitem{cmos}
\begin{barticle}
\bauthor{\bsnm{Chen}, \binits{C.}},
\bauthor{\bsnm{Wang}, \binits{Y.}},
\bauthor{\bsnm{Xu}, \binits{Y.}},
\bauthor{\bsnm{Zhao}, \binits{Z.}},
\bauthor{\bsnm{Qiu}, \binits{H.}},
\bauthor{\bsnm{Hou}, \binits{D.}},
\bauthor{\bsnm{Yang}, \binits{X.}},
\bauthor{\bsnm{Ma}, \binits{J.}},
\bauthor{\bsnm{Chen}, \binits{Y.}},
\bauthor{\bsnm{Zhao}, \binits{Y.}},
\bauthor{\bsnm{Liu}, \binits{H.}},
\bauthor{\bsnm{Zhao}, \binits{X.}},
\bauthor{\bsnm{Zhu}, \binits{Y.}}:
\batitle{{Performance of a focal plane detector for soft {X}-ray imaging
  spectroscopy based on back-illuminated sCMOS}}.
\bjtitle{Nuclear Inst. and Methods in Physics Research, A}
\bvolume{1030},
\bfpage{16646501}--\blpage{16646508}
(\byear{2022})
\end{barticle}
\endbibitem

%%% 21
\bibitem{contamination}
\begin{barticle}
\bauthor{\bsnm{O'Dell}, \binits{S.L.}},
\bauthor{\bsnm{Elsner}, \binits{R.F.}},
\bauthor{\bsnm{Oosterbroek}, \binits{T.}}:
\batitle{{Effects of contamination upon the performance of x-ray telescopes}}.
\bjtitle{Proceedings of SPIE - The International Society for Optical
  Engineering}
\bvolume{7732},
\bfpage{77322}--\blpage{017732216}
(\byear{2010})
\end{barticle}
\endbibitem

%%% 22
\bibitem{STMPanterTest}
\begin{barticle}
\bauthor{\bsnm{Miranda}, \binits{B.}},
\bauthor{\bsnm{Vadim}, \binits{B.}},
\bauthor{\bsnm{Peter}, \binits{F.}},
\bauthor{\bsnm{Gisela}, \binits{H.}},
\bauthor{\bsnm{Andreas}, \binits{L.}},
\bauthor{\bsnm{Giuseppe}, \binits{V.}},
\bauthor{\bsnm{Dervis}, \binits{V.}},
\bauthor{\bsnm{Yong}, \binits{C.}}:
\batitle{{X-ray testing of the Einstein Probe follow-up x-ray telescope STM at
  MPE's PANTER facility}}.
\bjtitle{SPEIE}
\bvolume{11444},
\bfpage{1144457}--\blpage{1114445712}
(\byear{2020})
\end{barticle}
\endbibitem

%%% 23
\bibitem{FXTchen}
\begin{barticle}
\bauthor{\bsnm{Chen}, \binits{Y.}},
\bauthor{\bsnm{Cui}, \binits{W.}},
\bauthor{\bsnm{Han}, \binits{D.}},
\bauthor{\bsnm{Wang}, \binits{J.}},
\bauthor{\bsnm{Yang}, \binits{Y.}},
\bauthor{\bsnm{Wang}, \binits{Y.}},
\bauthor{\bsnm{Ma}, \binits{J.}},
\bauthor{\bsnm{Xu}, \binits{Y.}},
\bauthor{\bsnm{Lu}, \binits{F.}},
\bauthor{\bsnm{Chen}, \binits{H.}},
\bauthor{\bsnm{Tang}, \binits{Q.}},
\bauthor{\bsnm{Yuan}, \binits{W.}},
\bauthor{\bsnm{Friedrich}, \binits{P.}},
\bauthor{\bsnm{Meidinger}, \binits{N.}},
\bauthor{\bsnm{Keil}, \binits{I.}},
\bauthor{\bsnm{Burwitz}, \binits{V.}},
\bauthor{\bsnm{Eder}, \binits{J.}},
\bauthor{\bsnm{Hartmann}, \binits{K.}},
\bauthor{\bsnm{Zhang}, \binits{J.}}:
\batitle{{Status of the follow-up x-ray telescope onboard the Einstein Probe
  satellite}}.
\bjtitle{Proc of SPIE}
\bvolume{11444},
\bfpage{885}--\blpage{893}
(\byear{2020})
\end{barticle}
\endbibitem

%%% 24
\bibitem{XMM-Newton}
\begin{barticle}
\bauthor{\bsnm{Gondoin}, \binits{P.}},
\bauthor{\bsnm{Aschenbach}, \binits{B.}},
\bauthor{\bsnm{Beijersbergen}, \binits{M.}},
\bauthor{\bsnm{Egger}, \binits{R.}},
\bauthor{\bsnm{Jansen}, \binits{F.}},
\bauthor{\bsnm{Stockman}, \binits{Y.}},
\bauthor{\bsnm{Tock}, \binits{J.P.}}:
\batitle{{Calibration of the first XMM Flight Mirror Module II - Effective
  Area}}.
\bjtitle{Proc of SPIE}
\bvolume{3444},
\bfpage{290}--\blpage{301}
(\byear{1998})
\end{barticle}
\endbibitem

\end{thebibliography}

\end{document}